\documentclass[11pt,A4,aps,showkeys,showpacs]{revtex4} 

\usepackage{times,fancyhdr}
\usepackage{epsfig}
\usepackage{epstopdf}   
\usepackage[latin1]{inputenc}
\usepackage[T1]{fontenc}

\begin{document}
\title{
 Majorana-like fermion physics:
 Emergence of topologically protected vortical states in graphene
 interacting with an electromagnetic field
}
\author{H. V. Grushevskaya}
\email{grushevskaja@bsu.by}
\affiliation{Belarusian State University, 4 Nezavisimosti Ave., 220030 Minsk,
Belarus}
\author{George Krylov}
\email{krylov@bsu.by}
\affiliation{Belarusian State University, 4
Nezavisimosti Ave., 220030 Minsk, Belarus}

\begin{abstract}
Within the framework of a quasi-relativistic model of graphene
that admits topologically nontrivial Majorana-like quasiparticle
excitations, appearance of such vortex states in the frequency
dependencies of the complex dielectric permittivity of the system
subjected to an external electromagnetic field has been examined.
The vortex graphene states possess topological charges (flavours ) being
nonzero Zak phases. Interaction
effects of  Majorana-like modes have been qualitatively
related to the formation of Fano resonances in the optical
response. The constructed topological model of graphene may be
considered as a toy model  of three-flavour mass-neutrino oscillations.
\end{abstract}

\pacs{72.80.Vp, 12.60.-i}

\keywords{Majorana-like fermion,
graphene, topologically protected vortical state, electromagnetic
field.}

\date{}
\maketitle


\pagestyle{fancy} \fancyhead{} \fancyhead[EC]{H. V. Grushevskaya
and G.~G. Krylov} \fancyhead[EL,OR]{\thepage}
\fancyhead[OC]{Majorana-like fermion physics: Emergence of
topologically protected vortical states in graphene interacting
with an electromagnetic field} \fancyfoot{}
\renewcommand\headrulewidth{0.5pt}

-------------------------------
\section{Introduction}
%
Majorana physics of high-energy interactions is currently actively
searched for after obtained experimental evidence of the existence
of non-zero masses in left-handed (LH) neutrinos. The neutrino
masses change periodically (see \cite{Esteban2020} and references therein).
%
%
There exist several scenarios for the origin of the three-flavour
neutrino oscillations. The simplest explanation, which extends the
Standard Model (SM) of Particle Physics, suggests a Majorana
origin for neutrinos.
%
%
The Majorana fields perform neutrino-electromagnetic-plasma
coupling through the Higgs exchange in Early Universe. The right-handed (RH) field
$\chi$ interacts with both a RH SM quark $q_R$ and
a mediator to SM gauge field $\tilde{q}$ via a Yukawa interaction
proportional to $ \tilde{q} \overline{q}_R \chi $
\cite{Escudero2019,Binder2023}.
%
%
The scenario is based on the fact that the elusive RH particle $\chi$ is free due to the suppression of
its interaction with other particles by CP symmetry. Such
elementary particles are called sterile heavy leptons. The
three-flavour neutrino oscillations are evidence of the fact that
the Majorana mass term should disappear. It is assumed that the
vanishing of the non-zero Majorana mass term with the subsequent
transformation of the LH Majorana particles into Weyl ones occurs
during the decoupling.
%
%
%
Such neutrinos are called active ones. Instead of describing the
decoupling stage, the mixing states of the LH active neutrinos
with the sterile lepton are phenomenologically calculated by
giving the observed pattern of masses and mixing in order to form
a combination of both the sterile RH particle and the massless LH
Weyl neutrinos \cite{Esteban2019}.
%
%
The difficulty with this approach is that, along with the Majorana
mass, the Yukawa interaction and, correspondingly, the sterile
heavy leptons should  disappear as well. The mechanism for the
periodic alternation of the Majorana and Dirac mass spectra,
referred to as the seesaw mechanism, is considered as one of the
SM big puzzles.
%
%
In another scenario, it is proposed to explain the neutrino mass spectrum
through an electroweak-symmetry-breaking RH partner to the left-handed Weyl neutrino (see
\cite{Vries2025} and references therein).
%
%
Having acquired a mass through Yukawa coupling, the sterile RH
neutrino remains free due to CP symmetry, but periodically mixes
with active LH massless Weyl neutrinos, providing them an apparent
mass. However, the issue of Majorana physics inevitably arises
after the addition of the right-handed Weyl neutrinos. The
introduction of the RH neutrino is a model complication, but any
complication can be viewed as a drawback, added to the same
drawbacks as in the first scenario.

Majorana processes 
such as neutrinoless double beta decays (see \cite{Chaturvedi2024} and references therein) are very rare
events because of 
the TeV energy range.
%
%
Until now, improvements to detector capabilities for rare decays
have been focused on neutrino double-beta decays, which meet much
more frequently than Majorana neutrinoless double-beta decays.
Attempts to observe ultra-rare Majorana events are beyond the
capabilities of current detectors \cite{Adams2026}.
%
Active  neutrinos can oscillate into sterile neutrinos \cite{Acero-et-al2024}.
But, all unsuccessful search for the appearing sterile neutrino \cite{Gagliardi-et-al2025}
points out that scenarios beyond the active-sterile mixing framework may be required.

Thus, in the absence of experimental power, the confirmation of
theoretical predictions of Majorana physics is problematic.
%
%

Analogues of hypothetical elementary Majorana and Dirac-Weyl particles are invoked
to explain topologically nontrivial quantum phases of strongly correlated
many-body systems. Zero-dimensional Majorana modes of one-dimensional Kitaev
chains are observed experimentally as the emergence of nonzero electron density
at the ends of the atomic chains.
%
%
Two-dimensional (2D) Majorana modes can be represented as
two-dimensional vortices with a topological defect inside.
However, theoretical predictions have significant discrepancies
with  experimentally observed optical and electrical properties of
topological unconventional superconductors, and just as in high
energy physics, the 2D Majorana modes remain elusive
single-particle excitations in solid-state physics.
%
%

Generally speaking, massless fermions are chiral ones. After they
acquire a mass, a revision of the chirality of the processes and
subsequent search for the manifestations of chirality anomalies
are required. Graphene is a strongly correlated many-electron
system that is testified by the existence of correlation-induced
spin-charge separation for graphene states \cite{Ren2023}.
%
%

In this paper, we study a Majorana-like fermionic graphene
model with topological subgap states and non-Abelian statistics.
We will show that the Majorana-like fermions carry topological charge, and the law
of conservation of topological charge allows only for interactions
between zero-energy Majorana modes. However, this Majorana-like
interaction manifests itself in a chiral anomaly. We demonstrate
that the gapping of the Majorana electron band structure of
graphene in magnetic fields leads to the expulsion of three
Majorana-like modes to the first Landau level.
%
This chiral anomaly is observed as Fano resonances of the complex  graphene
dielectric function at excitation near the Dirac valleys $K(K')$
of the graphene Brillouin zone; when the Majorana mass term disappears, the response to electromagnetic excitation
 of the electron density in flat regions near the $M$ points of the graphene
  Brillouin zone oscillates. We aim to interpret this behavior as a seesaw
  mechanism within the framework of a toy Majorana-like fermion model
   of high-energy physics with three-flavour neutrino oscillations.

\section{Majorana-like physics of graphene fermions in electromagnetic fields  }

\begin{figure}[htbp]
\centering
\hspace{0.3cm}(a)\hspace{4cm} (c) 
\\
\includegraphics[width=4.0cm]{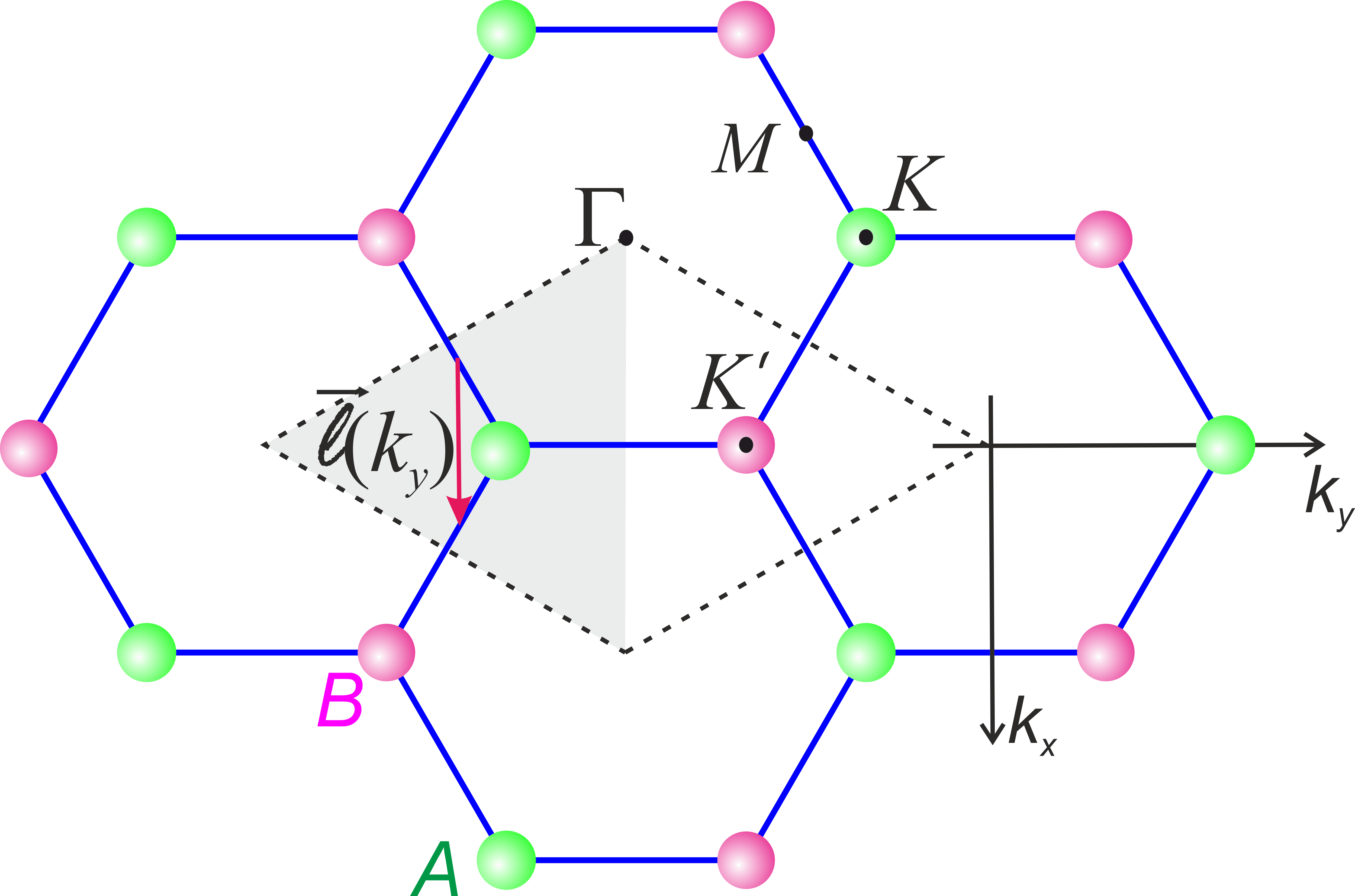}\hspace{0mm}
\includegraphics[width=7.5cm,angle=0]{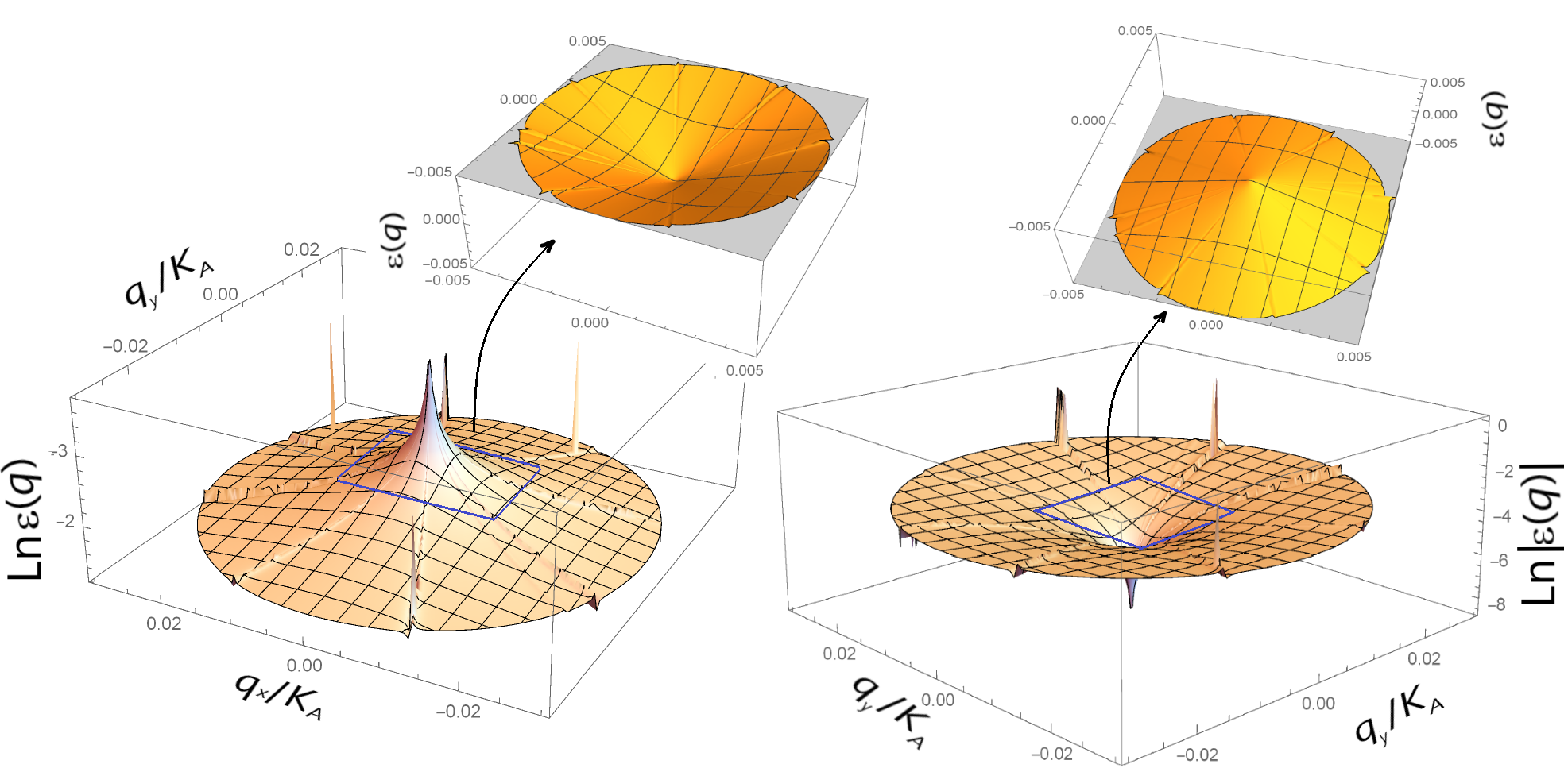}%
\\
(b)\\
\includegraphics[width=6.cm,angle=0]{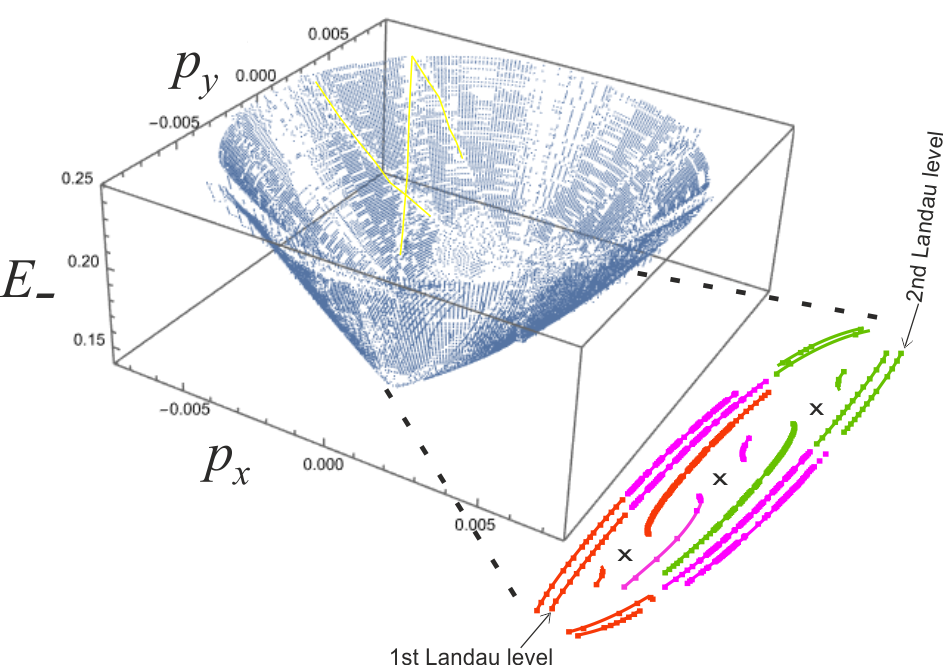}\
\includegraphics[width=6.cm,angle=0]{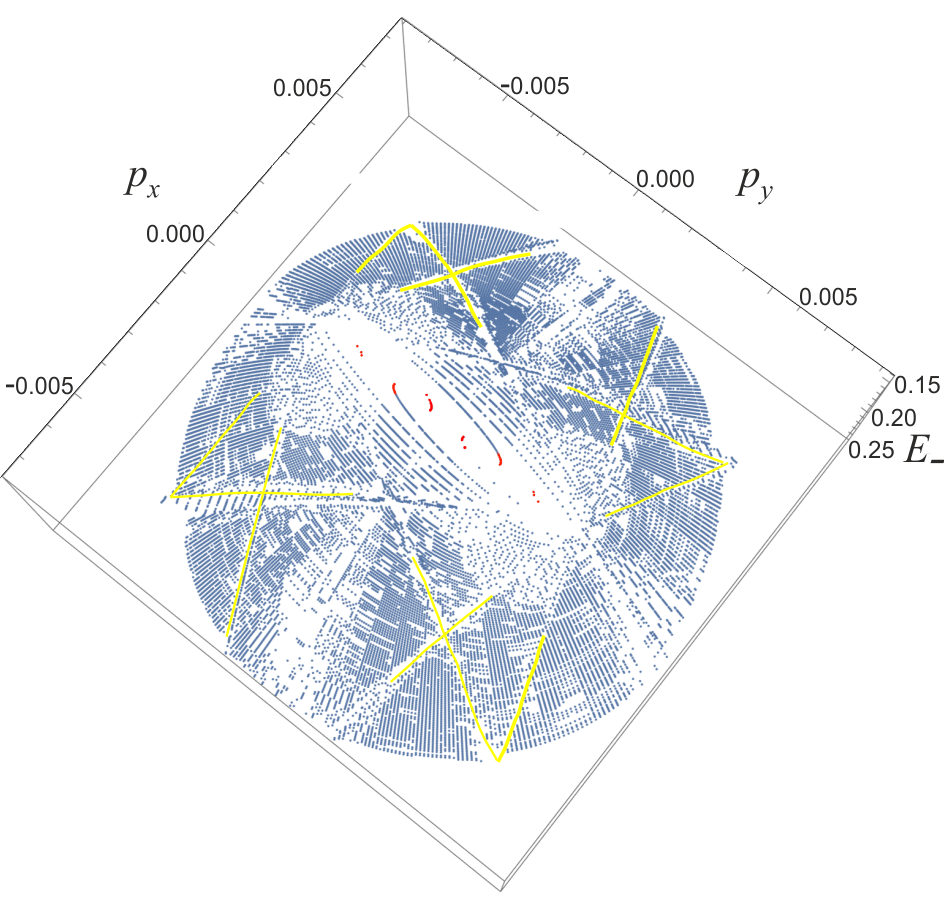}
\caption{(a) Graphene hexagonal lattice of Brillouin zone.
A change $\vec \gamma \left[\vec l(k_y)\right]$ in the flux of non-Abelian gauge field
along the paths $\vec l (k_y)$ defines Wilson loops. A rhombic
first Brillouin zone (BZ) of the honeycomb lattice consisting of
two triangular sublattices $A$ (green) and $B$ (red) is labeled
with a dashed line. High-symmetry points are $\Gamma,\ K(K') , \
M$. Occupied half-BZ is gray shaded. A reference point of
coordinates $(k_x,k_y)$ is in $\Gamma$.
(b)~A  valence (conduction) band of Majorana fermions in
graphene subjected to an electromagnetic field: (left) side view 
and (right)  top view . 
The yellow lines provide a guide to the eye for the 4  conical and 4
pseudo-Dirac one-dimensional forbidden zones. The excited Majorana modes are painted in red color.
The figure, bottom 
and Inset in it 
illustrate the cyclotron gap (the first Landau level) appearing due to the Landau quantization and a splitting of
the  shifted Majorana zero-energy mode entering the quantized spectrum of graphene.
The inset shows the bottom of the perturbed band up to and
including the 2nd Landau level; the three discontinuities  in the anomalous
dispersion energy levels curves are designated by ``crosses''. The
three  excited Majorana modes with the corresponding two Landau
levels are colored in orange, magenta, and green. (c) The graphene
bandstructure non-perturbed by external field action.}
\label{Bandstructure}
\end{figure}

Graphene is a hexagonal layer of carbon atoms with one atom thick.
Its two electrons per primitive rhombic unit cell are located
on the trigonal sublattices $A$ and $B$ of the hexagonal lattice.
A Landau level with energy at
a point or valley $K(K')$ named as the Dirac point of graphene Brillouin zone is half-filled in neutral graphene.
Fig.~\ref{Bandstructure}a shows the graphene Brillouin zone.
Therefore, the wave functions 
$\psi_{AB}$ and $\psi^*_{BA}$ of  graphene charge carriers are attributed to the
wave function
of graphene sublattices $A$ and $B$, respectively; here $*$ is complex conjugation. The graphene wave function
$\left( \psi_{AB}, \psi^*_{BA}\right)^T$  is the wave of electron--hole pair being resident in graphene,
where $T$ is the transposition operation
.
Since the electron--hole pair is simultaneously its own antiparticle,
a model with fermions similar to Majorana ones is necessary to describe graphene.
Such a quasi-relativistic graphene model has been derived in \cite{Taylor2016}
as a consequent account of the effect of relativistic exchange interactions. The model
is grounded on truly secondary quantized relativistic consideration of the problem within
the known Dirac--Hartree--Fock
self-consistent field approximation. In subsequent publications
\cite{our-symmetry2020,myNPCS18-2015,NPCS18-2015GrushevskayaKrylovGaisyonokSerow}
it has been established that the model admits a form as Majorana-like system of equations
as well as two-dimensional Dirac-like equation with
an additional ``Majorana-force correction'' term \cite{our-symmetry2020}.

The  Hamiltonian of  Majorana-like fermions in an external electromagnetic field
  reads \cite{8-braid,our-symmetry2020,myJNPCS2025poles}
\begin{eqnarray}
&\left[c\vec \sigma_{2D}^{BA}\cdot \left(\vec p_{AB}- {e\over c}\vec A\right)
- \widetilde{M}_{AB}
\left(\vec p_{AB}- {e\over c}\vec A\right) \right]\left| \psi_{AB}\right\rangle =
i c{\partial \over \partial t}\left| \psi^*_{BA}\right \rangle , \label{Majorana-bispinor01}\\
&\left[c \vec \sigma_{2D}^{AB}\cdot \left( \vec p\,^*_{BA} + {e\over c}\vec A \right)
-\left( \widetilde{M}_{BA}
\left(\vec p_{BA}- {e\over c}\vec A\right) \right)^*\right] \left|\psi_{BA}\right\rangle
  = - ic {\partial \over \partial t}\left|\psi^*_{AB}\right \rangle . \quad
\label{Majorana-bispinor02}
\end{eqnarray}
%
Here $\widetilde{M}_{AB} $ ($\widetilde{M}_{BA}$) is an unconventional
Majorana-like mass  term for a  quasiparticle in the sublattice
$A(B)$; $\vec \sigma_{2D}^{AB}= \Sigma_{BA} \vec \sigma_{2D}\Sigma_{AB}^{-1}$,
$\vec \sigma_{2D}^{BA}= \Sigma_{AB} \vec \sigma_{2D}\Sigma_{AB}^{-1}$,
$\vec \sigma_{2D} =\{\sigma_x, \sigma_y\}$ is the 2D vector of the Pauli
matrixes;
$\vec p_{2D}^{AB}= \Sigma_{AB}\ \vec p\ \Sigma_{AB}^{-1}$,
$\vec p_{2D}^{BA}= \Sigma_{BA}\ \vec p \ \Sigma_{BA}^{-1}$, $\vec p=\{p_x, p_y\}$ is the 2D momentum operator,
 $\Sigma_{AB}$ and $\Sigma_{BA}$ are
relativistic quantum-exchange operators for sublattices $A, B$
respectively; $\vec A$ is a vector-potential of electromagnetic field, $c$ is the speed of light,
$e$ is the electron charge
.
The Majorana-like mass terms $\widetilde{M}_{BA}$ and  $\widetilde{M}_{AB}$ are determined as
\begin{eqnarray}
\widetilde{M}_{BA}=\alpha i^2\Sigma_{BA}\Sigma_{AB}- \hbar c
(\vec \sigma^{AB} \cdot \vec p^{BA} )  \vec \sigma^{AB} \cdot
\left(\vec K_B^{BA} + \vec K_A^{BA}\right),\label{orbital-valley-currents-coupling1}\\
\widetilde{M}_{AB}=\alpha i^2\Sigma_{AB}\Sigma_{BA}
- \hbar c
(\vec \sigma^{BA} \cdot \vec p^{AB} )  \vec \sigma^{BA} \cdot
\left(\vec K_B^{AB} + \vec K_A^{AB}\right)
\label{orbital-valley-currents-coupling2}
\end{eqnarray}
where $\alpha$ is an arbitrary constant,
$\vec K_A^{BA}=\Sigma_{BA} \vec K_A \Sigma_{BA}^{-1}$ ($\vec K_B^{BA}=\Sigma_{BA} \vec K_B \Sigma_{BA}^{-1}$),
$\vec K_A^{AB}=\Sigma_{AB} \vec K_A \Sigma_{AB}^{-1}$ ($\vec K_B^{AB}=\Sigma_{AB} \vec K_B \Sigma_{AB}^{-1}$),
$\vec K_A$ and $\vec K_B$ denote the valleys $\vec K$ and $\vec K'$, $\hbar=h/(2\pi)$, $h$ is the Planck constant.
You can see that $\widetilde{M}_{BA} (\widetilde{M}_{AB})$
is comprised by  the two terms being the proper Majorana mass term which is proportional to
$\Sigma_{BA}\Sigma_{AB}$ ($\Sigma_{AB}\Sigma_{BA}$) and the  coupling
\begin{eqnarray}
V_{o-v}=- \hbar c (\vec \sigma^{AB} \cdot \vec p^{BA} )  \vec \sigma^{AB} \cdot
\left(\vec K_B^{BA} + \vec K_A^{BA}\right) \label{spin-valley-coupling}
\end{eqnarray}
 between orbital and valley currents.
In what follows we vary 
the parameter $\alpha$ and omit the symbol $2D$
in designations of the transformed vector of $2D$ Pauli matrixes.
The presence of  the coupling between orbital and valley current means that the
problem (
\ref{Majorana-bispinor01})--(\ref{Majorana-bispinor02}) is effectively many-body one due to account of
electron-hole correlations.

The relativistic exchange operator for the tight-binding approximation and accounting of
nearest lattice neighbors is given by its action on secondary quantized wave functions
on sublattices $A(B)$ of the system
\cite{Taylor2016,myNPCS18-2015,NPCS18-2015GrushevskayaKrylovGaisyonokSerow}
in the following form:
\begin{eqnarray}
\Sigma_{rel}^{x}\left(
\begin{array}{c}
\widehat {\chi } ^{\dagger}_{_{-\sigma_{_A}} }(\vec r) \\
\widehat {\chi }^\dagger _{\sigma_{_B}}(\vec r)
\end{array}
\right)\left|0,-\sigma \right> \left|0,\sigma \right>  \ \ \ \ \\ \nonumber
= \left(
\begin{array}{cc}
0&  \Sigma_{AB}
\\
\Sigma_{BA} & 0
\end{array}
\right)
\left(
\begin{array}{c}
\widehat {\chi }^{\dagger}_{-\sigma_{_A} } (\vec r) \\
\widehat {\chi} ^\dagger _{\sigma_{_B}}(\vec r)
\end{array}
\right)\left|0,-\sigma \right> \left|0,\sigma \right>
\label{exchange}
,  
\end{eqnarray}
with
\begin{eqnarray}
\Sigma_{AB} \widehat {\chi }^\dagger _{\sigma_{_B}}(\vec
r)\left|0,\sigma \right>
= \sum_{i=1}^{N_v\,N}\int { d \vec r_i} \widehat {\chi }^\dagger
_{\sigma_i{^B}}(\vec r)\left|0,\sigma \right> \ \ \\ \nonumber
\times  \Delta _{AB} \langle
0,-\sigma_i|{\widehat \chi}^\dag_{-\sigma_i^A} (\vec r_i) V(\vec
r_i -\vec r) {\widehat \chi}_{-\sigma_B}(\vec
r_i)|0,-\sigma_{i'}\rangle , \label{Sigma-AB}
\\ 
 \Sigma_{BA}
\widehat {\chi }^{\dagger}_{_{-\sigma_{_A}} } (\vec r)
\left|0,-\sigma \right>
=\sum_{i'=1}^{N_v\,N}\int { d \vec r_{i'}} \widehat {\chi
}^{\dagger}_{_{-\sigma_{i'}^A} } (\vec r) \left|0,-\sigma \right>
\\
\times \Delta _{BA} \langle 0,\sigma_{i'}|{\widehat
\chi}^\dag_{\sigma_{i'}^B} (\vec r_{i'}) V(\vec r_{i'} -\vec r)
{\widehat \chi}_{_{\sigma_A}}(\vec r_{i'})|0,\sigma_i\rangle . \ \
\label{Sigma-BA}
\end{eqnarray}
Here %
 $V(\vec r)$ is the three-dimensional (3D) Coulomb potential,  summation is perfor\-med on
 all lattice sites and number of electrons,
the interaction ($2\times 2$)-matrices $\Delta _{AB}$ and $\Delta_{BA}$
are gauge fields (or components of a gauge field).
Vector-potentials for these gauge fields are introduced by the phases 
$\alpha_{ 0}$ and $\alpha_{\pm, k}$, $k=1,\ 2,\ 3$ of
$\pi(\mbox{p}_z)$-electron wave functions $\psi_{\mbox{\small
p}_z}(\vec r)$ and $\psi_{\mbox{\small p}_z, \pm \vec
\delta_k}(\vec r)$ attributed to a given lattice site and its three nearest neighbors
(see details in \cite{NPCS18-2015GrushevskayaKrylovGaisyonokSerow}).
The definition
of these three non-Abelian gauge fields 
was stipulated by a requirement of reality of eigenvalues of the Hamiltonian
operator as gauge conditions. The operator of relativistic
exchange  gains an additional implicit $\vec p$-dependence upon
momentum in the case of non-zero values of gauge fields
\cite{8-braid}.

\section{Chirality anomalies and deconfinement of Majorana-like modes with Fano-resonance coupling}

\begin{figure*}[hbtp]
\begin{center}
(a) \hspace{7 mm}  \\
\includegraphics[width=7cm]{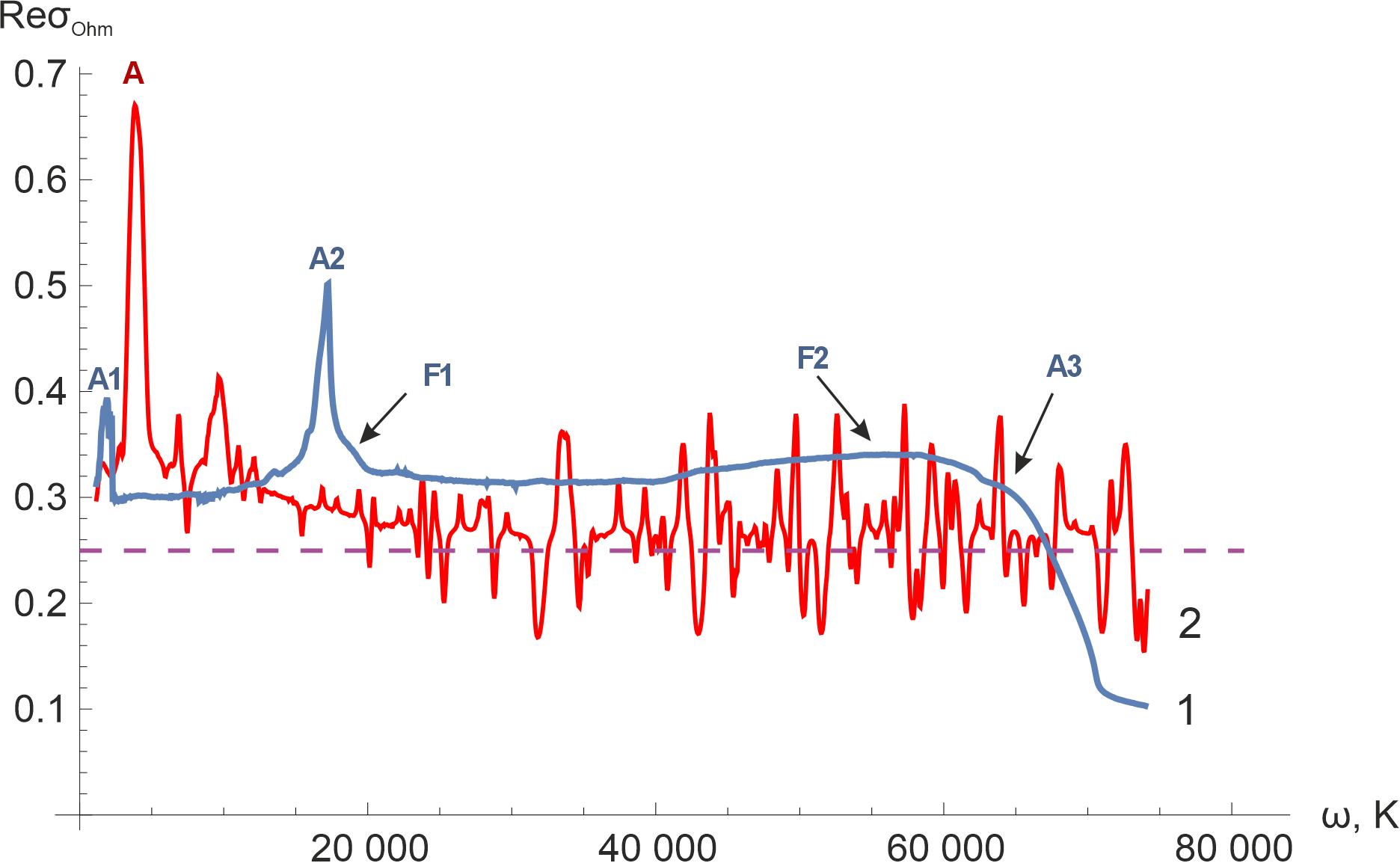}\
\includegraphics[width=7cm]{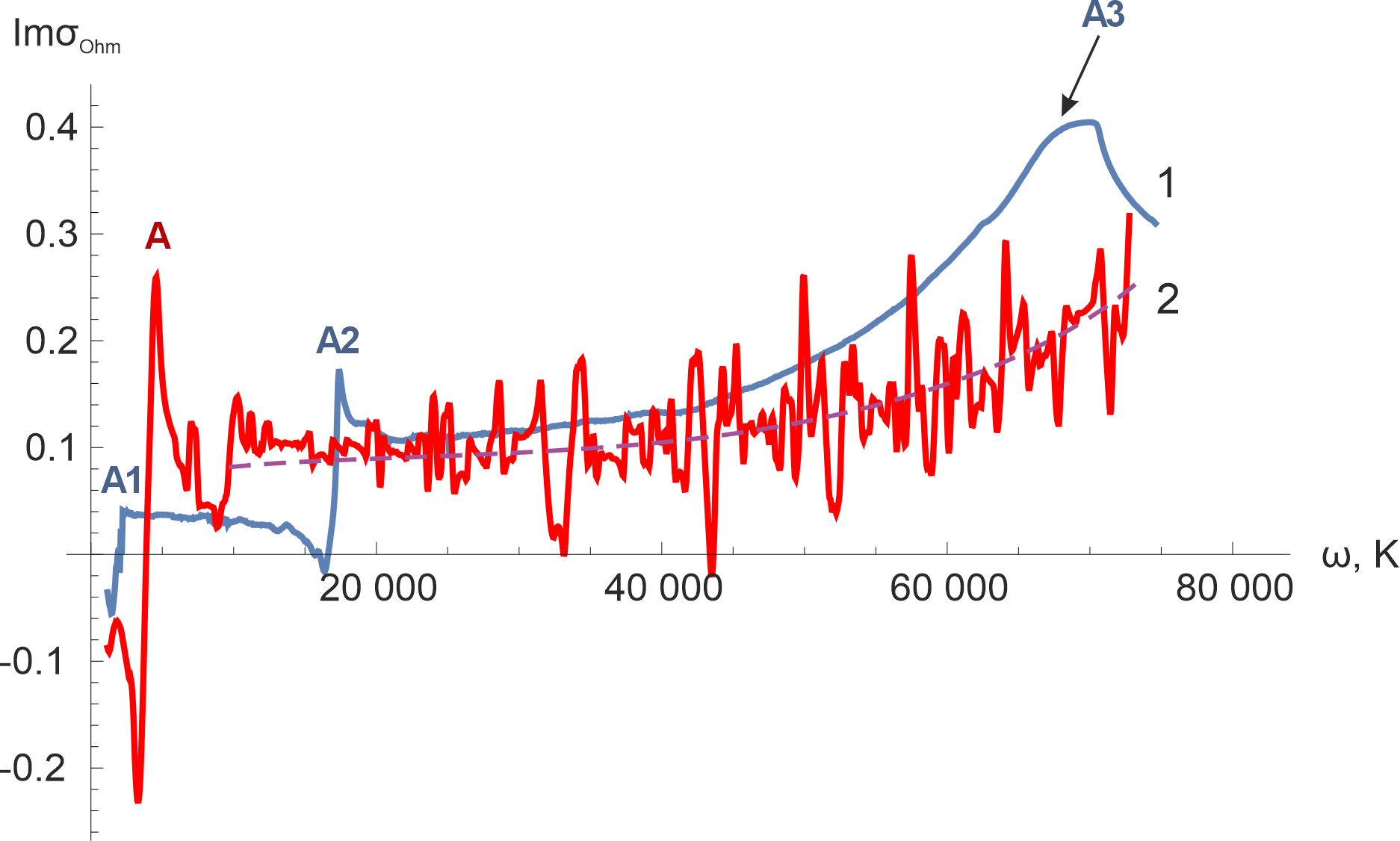}
\\ (b) \\
\includegraphics[width=7cm]{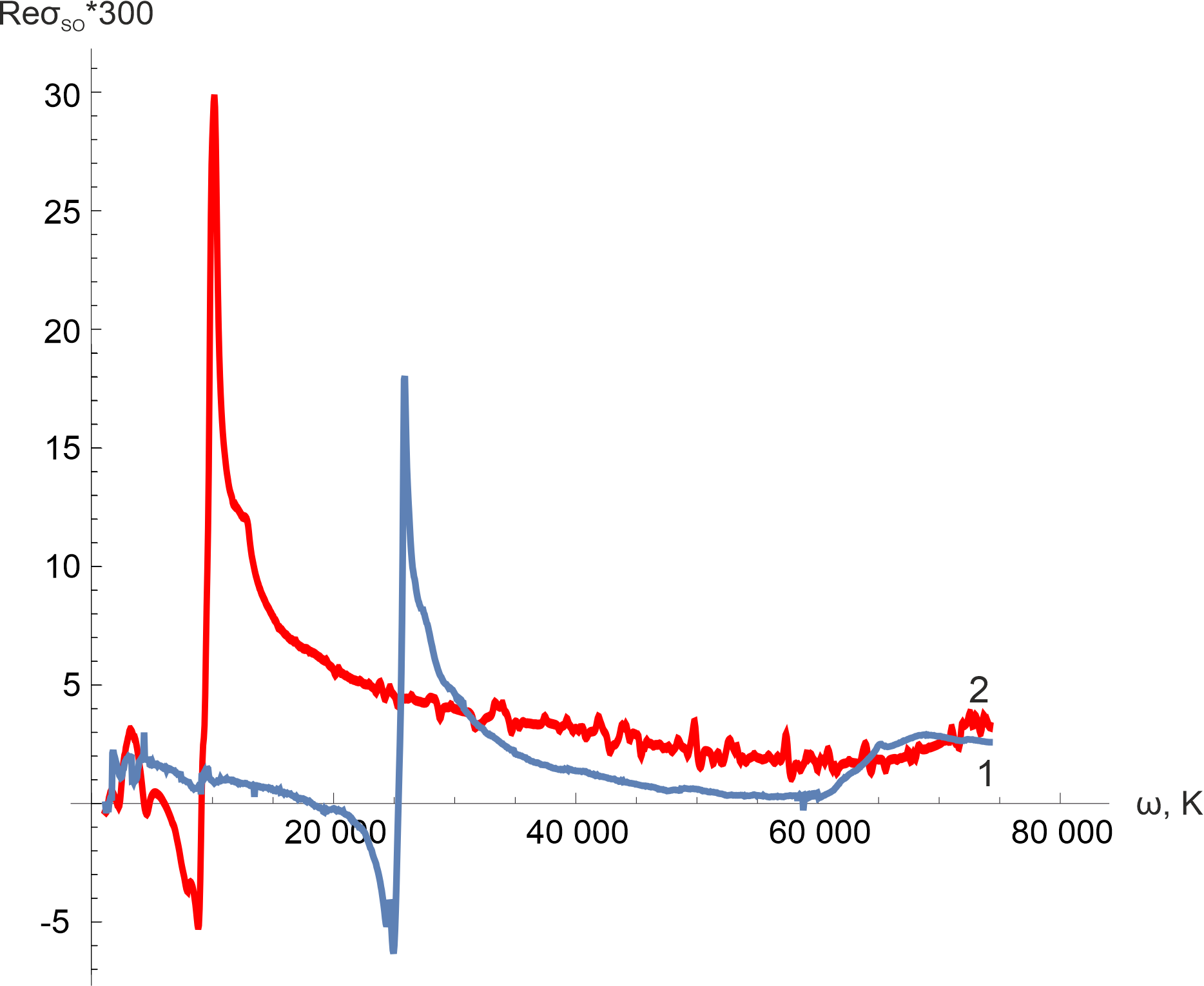}\hspace{0mm}
 \includegraphics[width=7cm]{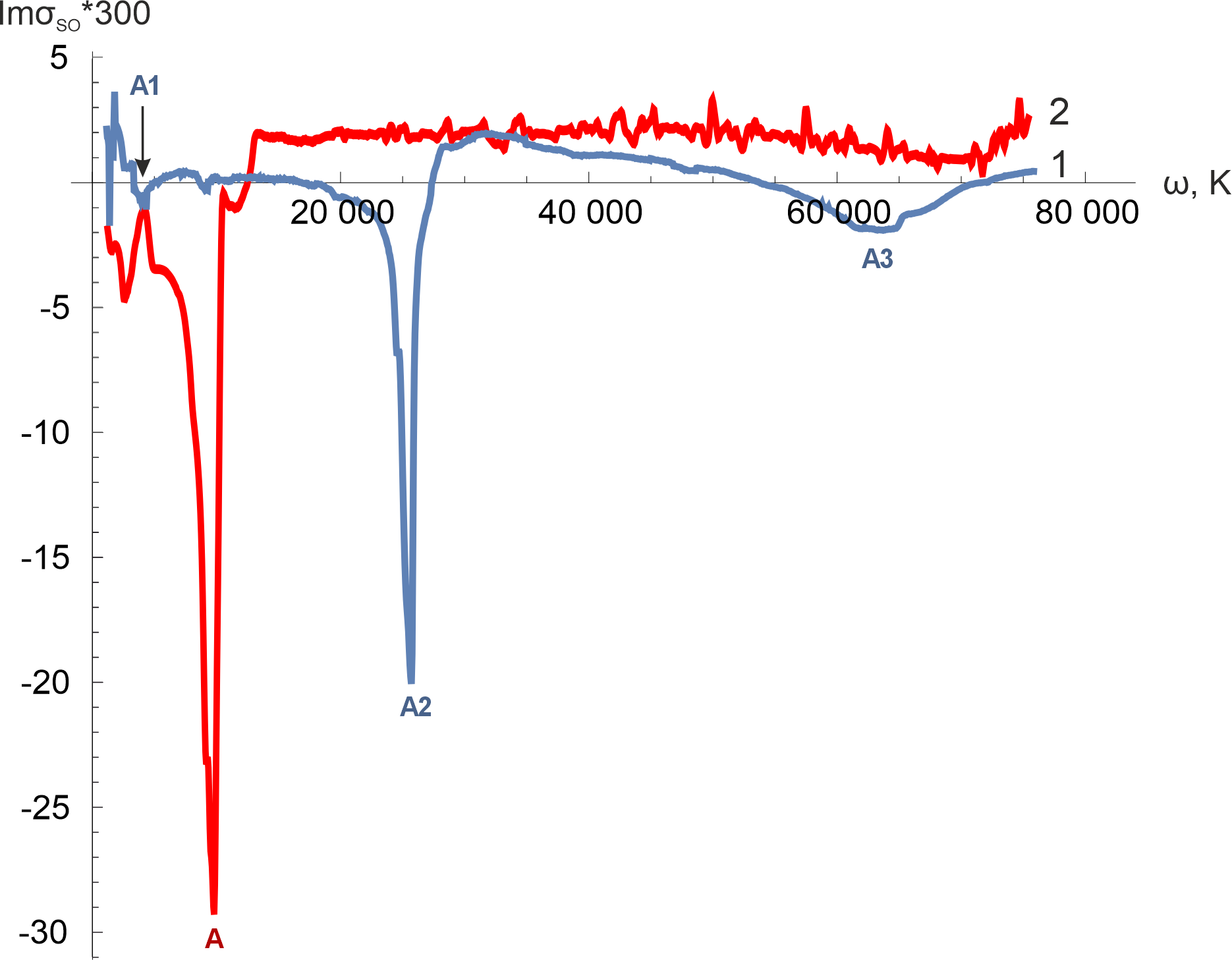}
\end{center}
\caption{Real (left) and imaginary  (right) 
of optical (a) and Hall (b) complex conductivity for graphene
Majorana-like fermion model with (blue curves ``1'') and
without (red curves ``2'') the mass term;
the dashed line is the optical conductivity of pseudo-Dirac graphene model.
``F1'' and ``F2'' denote Fano resonances,
``A'' and ``Ai'',$i=1,2,3$ denote degenerated and non-degenerated Majorana-like modes. }
 \label{Majorana-mode-excitation}
\end{figure*}

Let 
the multiplier
\begin{eqnarray}
m_{precession}\equiv \vec \sigma^{AB} \cdot \left( \vec K^{BA}_B+\vec K^{BA}_A \right) \label{m-precession}
\end{eqnarray}
entering the spin--valley-current coupling $V_{o-v}$ (\ref{spin-valley-coupling}) be considered as vanishing
\begin{eqnarray}
\vec \sigma^{AB} \cdot \left( \vec K^{BA}_{B}+\vec K^{BA}_{A} \right)= 0.
\end{eqnarray}
In magnetic fields, since discrete Landau energy levels arise the perturbed graphene bandstructure is gapped
(see Fig.~\ref{Bandstructure}b). Three excited Majorana-like modes
as the three anomalies of energy dispersion emerge in the vicinity of the valley
for the graphene model with a nonzero Majorana mass term
$\Sigma_{BA}\Sigma_{AB}\ (\Sigma_{AB}\Sigma_{BA})\neq 0$ (see Fig.~\ref{Bandstructure}b).
At zero magnetic fields, the unperturbed graphene band touches with
the  unperturbed graphene conduction  in the Dirac point as Fig.~\ref{Bandstructure}c shows.
%
These three Majorana modes 
are revealed  in the frequency dependence of the complex optical
conductivity  when $\Sigma_{BA}\Sigma_{AB}\ (\Sigma_{AB}\Sigma_{BA})\neq 0$.
The three anomalous dispersions   correspond to
 the three peaks  ''A1'', ``A2'', and ``A3'' 
 of the dielectric function (see Fig.~\ref{Majorana-mode-excitation}a).

The optical response of the Majorana-like graphene model deviates significantly
from the theoretically predicted constant behavior of its optical conductivity in the pseudo-Dirac theory.
In the ultraviolet range, a broad band with a maximum of $4.74$~eV ($5.5 \times 10^4$~K) is observed,
being asymmetric as a Fano resonance (see Fig.~\ref{Majorana-mode-excitation}a).
%
For massless graphene fermions of the Majorana type, these three peaks ``A1'', ``A2'', and ``A3'' corresponding to
the three anomalous dispersions  degenerate into the one peak labeled by ``A''.
Since the chiral anomalies vanish, the chirality of the model is restored, and therefore the optical
conductivity becomes, on average, equal to $0.25~G$, as for graphene fermions of pseudo-Dirac type.
Here $G/2$ is the quantum of minimal conductivity. It means that magnetic fields lift the degeneracy of the
Majorana-type bandstructure (see Fig.~\ref{Bandstructure}b).

For the topologically nontrivial pseudo-Majorana graphene model, a topological defect
is located at the Dirac point $K(K')$; charge carriers, bypassing this defect, acquire a nonzero phase called
the Zak phase as
\begin{eqnarray}
Q= -ie \oint_{C_{BZ}} {d\vec k \over (2\pi )^d}\cdot \left< u_{\vec k}(\vec r +\vec R(t_0+\epsilon))\right|
{ \partial\over \partial \vec k}\left| u_{\vec k}(\vec r +\vec R(t_0)) \right>,\
\epsilon \to 0
\end{eqnarray}
where $C_{BZ}$ is a path which is equivalent 
to a closed loop of $d$-dimensional Brillouin zone due
to crystal symmetry, $u_{\vec k}(\vec r +\vec R(t_0))$ is a Bloch function,
$t_0$, $t_0=0$ is a time.
The physical meaning of the Zak phase $Q$ consists in the emergence
of electric polarization (nonzero dipole moment) $\vec d = e \vec r$ \cite{Zak1989,Vanderbilt2018}
because the operator $i{\partial \over \partial \vec  k}$ is a position operator $\vec r$.
The transition dipole moment $\left< \vec K\right| \vec r\left| \vec K+\vec q\right>$, $\vec q\to 0$,
in the vicinity of the Dirac point is nonzero as the law of topological-charge conservation forbids its change.

%
The behavior of the real part of the complex Hall conductivity remains qualitatively the same,
whereas the imaginary contribution of the complex Hall conductivity to the graphene dielectric
 function has three plasmon modes (three regions of negative values)
 when $\Sigma_{BA}\Sigma_{AB}\ (\Sigma_{AB}\Sigma_{BA})\neq 0$, or one in the
 zero-Majorana-mass case (see Fig.~\ref{Majorana-mode-excitation}b).
%
These plasmons are oscillations of graphene quasiparticle excitations, the annihilation of which
is prevented by the polarization of the graphene electron density with the dipole moment
$\vec d$ per one electron--hole pair.
It means that bound charge carriers are always present in the graphene valleys.
Thus, topological defects effectively transfer a portion of charge
carriers from the valence band to the conduction band. In contrast
to the graphene Majorana-like fermions, pseudo-Dirac graphene charge carriers are absent at $K(K')$ in the
 low-frequency limit, $\omega_0 \to 0$.

The wave functions of an electron  and a hole, 
$\psi_e$ and $\psi_h$ of a non-topological electron--hole pair
and, correspondingly,  a wave $\Psi_{e-h}$ of the pair  acquire 
Pancharatnam--Berry phases
\cite{Huang-et-al2025,Griffiths-Schroeter}
 in a magnetic field $\vec B$ with the vector-potential $\vec A $ as
\begin{eqnarray}
\Psi_{e-h}(\vec A) = \psi_e \exp\left\{- {ie\over \hbar c}\oint_{C_e} d\vec r_e \cdot \vec A (\vec r_e )\right\}
\psi_h \exp\left\{- {ie\over \hbar c} \oint_{C_h} d\vec r_h  \cdot\vec A(\vec r_h )
\right\}\nonumber \\
= \Psi_{e-h}(0) e^{i(\gamma_e + \gamma_h)}
\end{eqnarray}
where $C_e$ ($C_h$)  is a loop electron (hole) path,
$\gamma_e$ ($\gamma_h$) is the electron (hole) Berry phase.
The wave functions of an electron and a hole, $\psi^T_e$ and $\psi^T_h$, of a topologically protected pair
acquire Berry phases, $\gamma^T_e$ and $\gamma^T_h$, opposite in sign and equal in magnitude,
$\gamma^T_e = - \gamma^T_h$, as the dipole moment $\vec d$ is attributed to the topological
electron--hole pair with a wave function $\Psi^T_{e-h}$.
The wave function $\Psi_{e-h}$ of the topologically protected electron--hole pair does not change, because
\begin{eqnarray}
\Psi^T_{e-h}(\vec A) = \psi^T_e \exp\left\{- {ie\over \hbar c}\oint_{C_{e,T}} d\vec r_e \cdot \vec A (\vec r_e )\right\}
\psi^T_h \exp\left\{- {ie\over \hbar c} \oint_{C_{h,T}} d\vec r_h  \cdot\vec A(\vec r_h )
\right\}\nonumber
\\ = \Psi^T_{e-h}(0) e^{i(\gamma^T_e + \gamma^T_h)}\equiv \Psi^T_{e-h}(0)
\label{hall-conductivity}
\end{eqnarray}
where $C_{e,T}$ ($C_{h,T}$) is a closed loop electron (hole) path around 
the topological defect, $C_{e,T}=C_{h,T}$, $\gamma_e$ ($\gamma_h$) is the  Berry phase for the electron (hole)
belonging to the topological pair.
Eq.~(\ref{hall-conductivity}) explains the conservative behavior of the real part of the Hall conductivity with
respect to the phase accumulation of electrons and holes composing topological electron--hole pairs in a magnetic
field $\vec B$ (see Fig.~\ref{Majorana-mode-excitation}b).
Thus, the Berry phase is not a topological invariant of the pseudo-Majorana graphene model.

Two Fano resonances (broad asymmetric bands of $\Re e \
\sigma_{Ohm}$), denoted as ``F1'' and ``F2'' in
Fig.~\ref{Majorana-mode-excitation}a, are evidence of the fact
that the Majorana modes are weakly coupled and interfere with a
fourth background process. The impact of the non-zero Majorana
mass term, $\tilde M$ gives
rise the Majorana-modes coupling as repulsion 
of the Majorana-like modes with same sign of their topological charges from each other 
and, hence, $\tilde M$ behaves as a certain interaction. 

Further, a mechanism of  the Majorana-modes flavour interaction
and an origin of the fourth background process are elaborated.

\section{Intervalley-precession coupling Majorana-like modes  as  a  topological toy model
 mechanism of neutrino oscillations}

\subsection{Emerging Majorana-like vortical states}

\begin{figure}[htbp]
\centering
\hspace{0.3cm}(a)\\
\includegraphics[width=8.0cm]{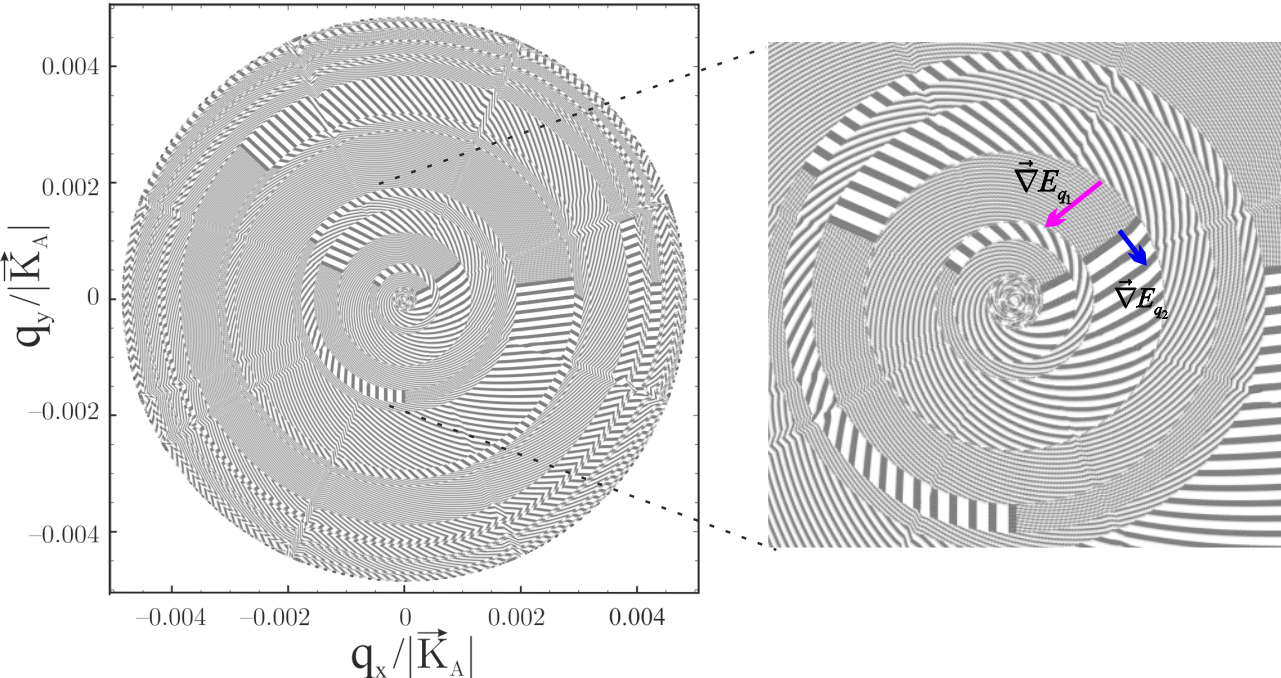}\\
\hspace{0.2cm}(b) \\
\includegraphics[width=4.0cm]{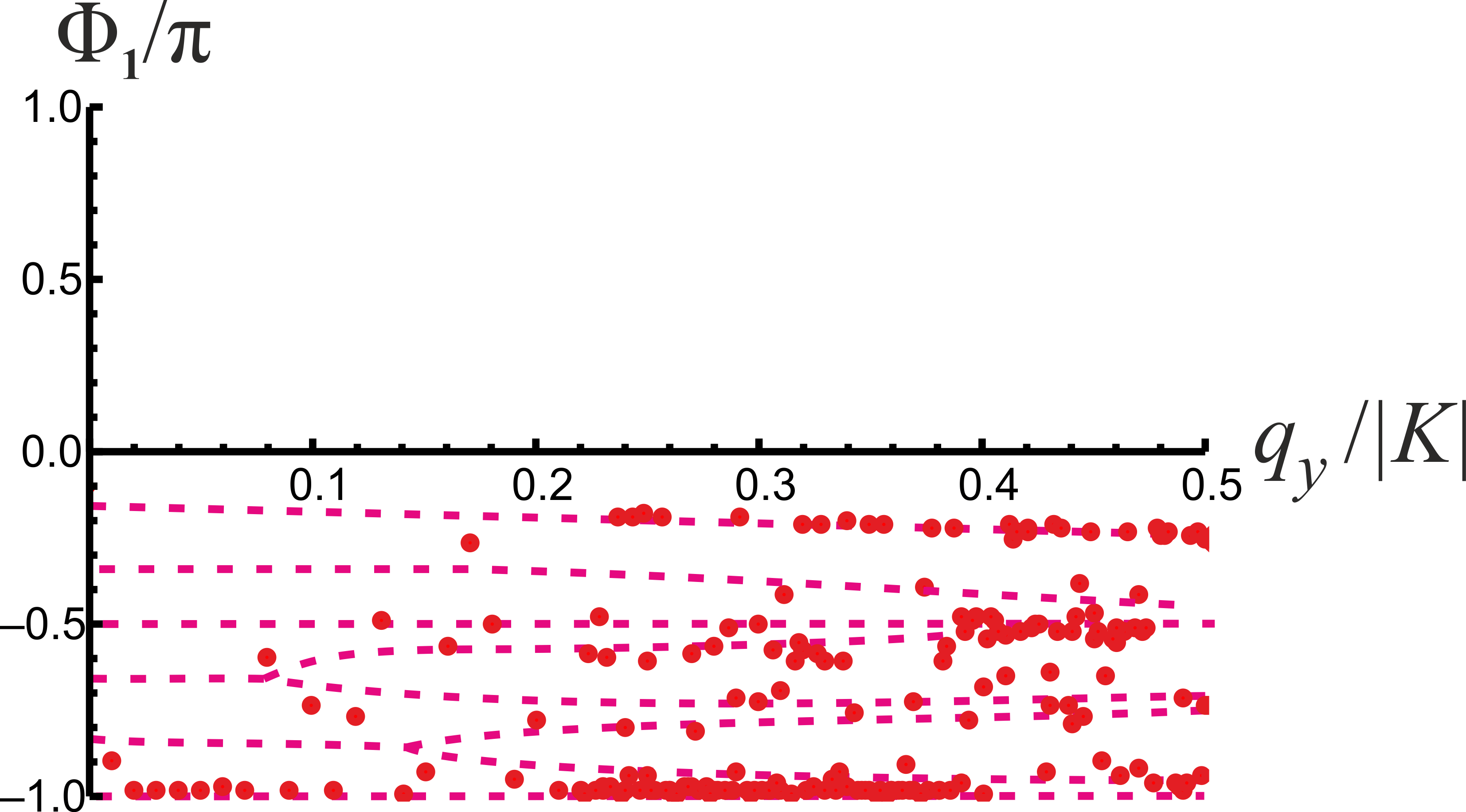} \hspace*{-1.0cm}
\includegraphics[width=4.0cm]{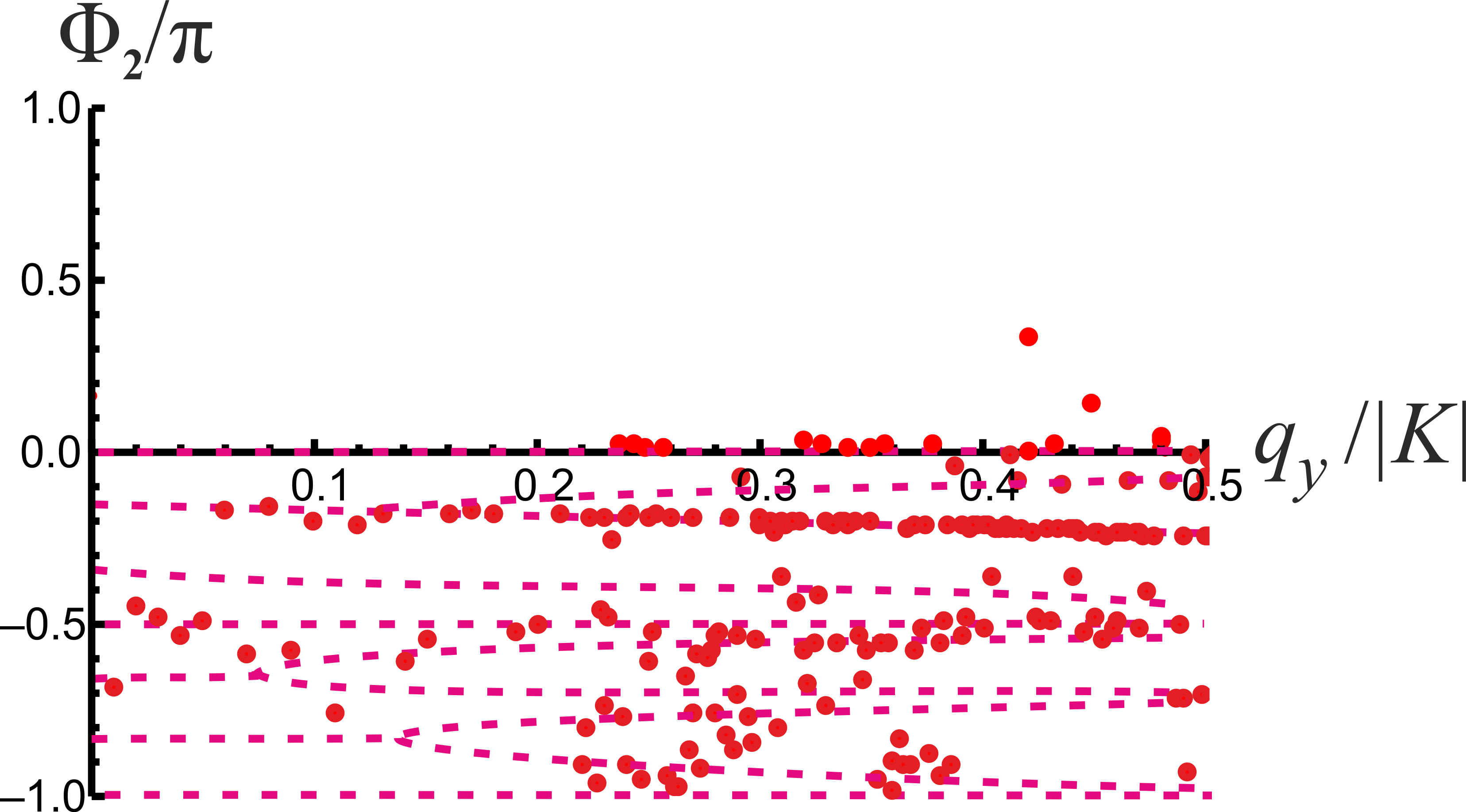} \hspace*{-1.0cm}
\includegraphics[width=4.0cm]{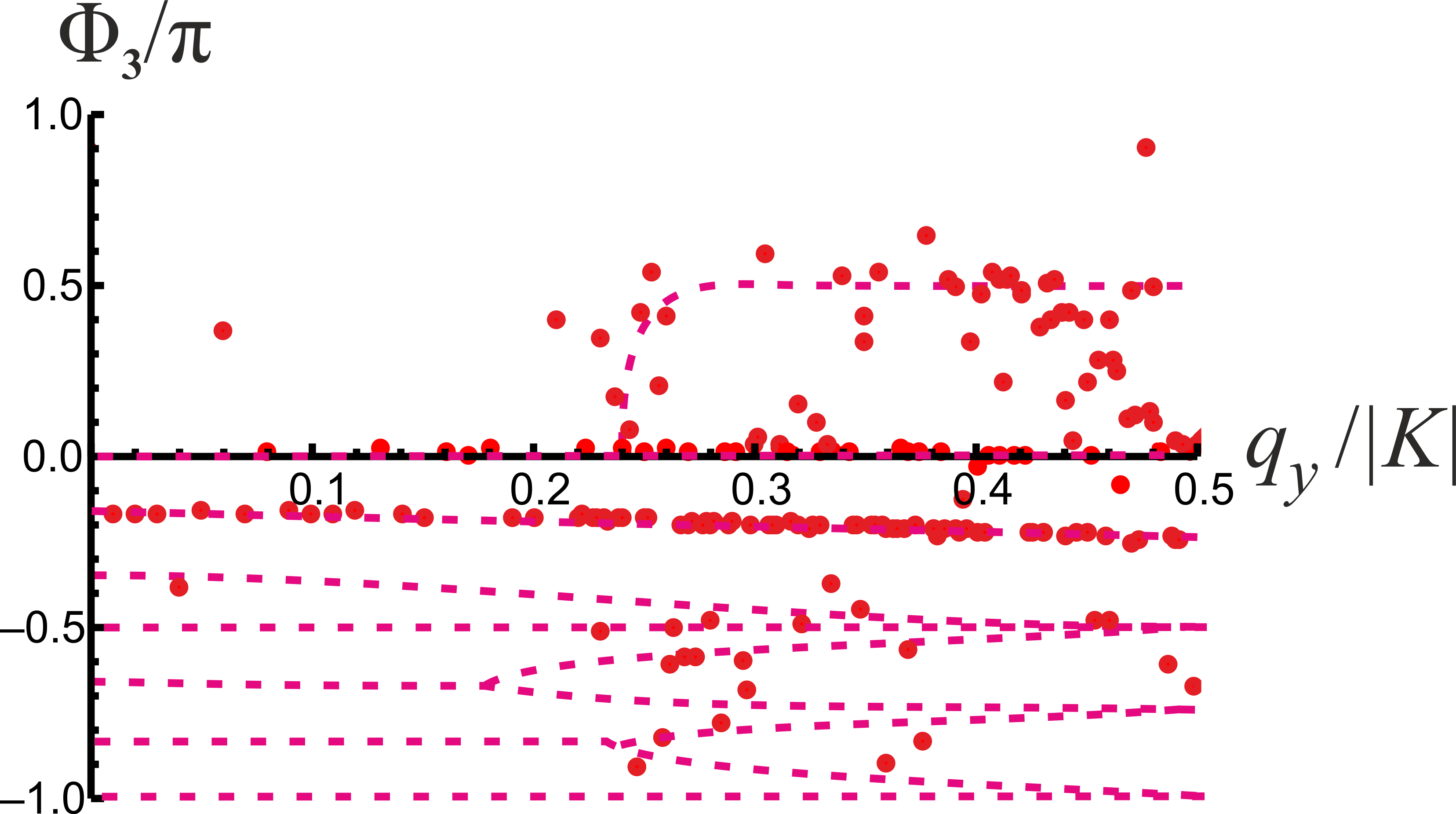}\hspace*{-1.0cm}
\includegraphics[width=4.0cm]{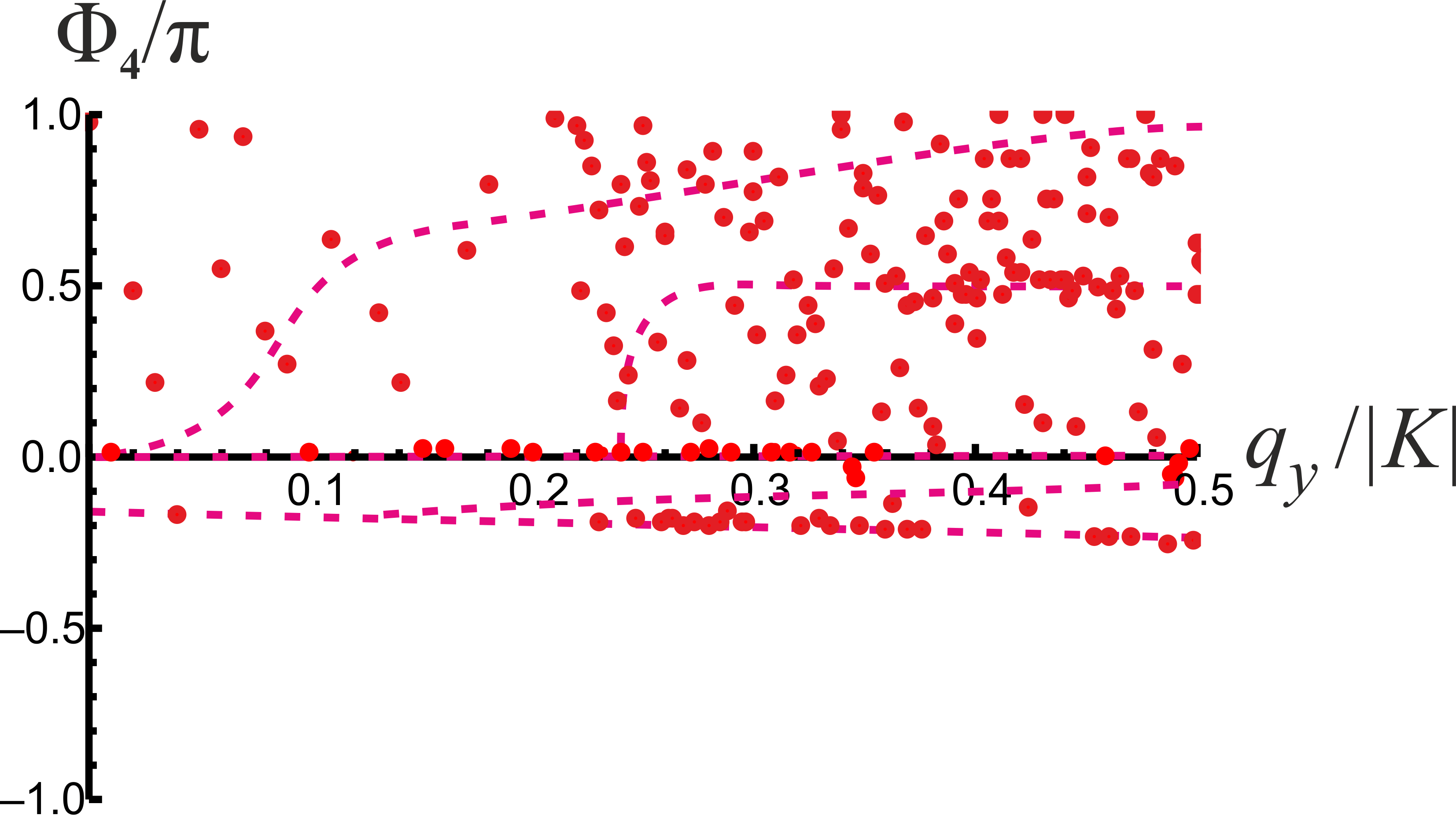}
\caption{(a) A spin--orbit texture of the bands on momentum scales $q/K=0.002$ in
contour plots  and a model  of intervalley-precession coupling Majorana-like modes  (in
inset to   (a)).
(b)~Non-Abelian phases $\Phi_1,\ldots,\Phi_4$
of the Wilson-loop eigenvalues in the units of $\pi$ at non-zero gauge fields. }
\label{fig1}
\end{figure}

In the absent of the electromagnetic field,
defects as core of vortexes 
remain in the Dirac point.
As one can see in Fig.~\ref{fig1}a, the contour
plots of the graphene energy electron valent (conduction) and hole
conduction (valent) bands are vortexed in the neighborhood of
$K(K')$ and the feathering of the vortex consists of an infinitely
many vortices of varying widths, called vortex sleeves.
%
Each vortex sleeve is located in a limited range of wave vector
values. Such discreteness is an attribute of quantization of the vortex sleeve.
Due to the law of conservation of angular momentum, both left- and
right-twisting vortices are generated.
Due to the law of conservation of the helicity of massless charge
carriers, the replacement $\vec k \to - \vec k$ produces a vortex
LH hole (electron) conductivity band from the RH electron (hole) valent band.

The vortical graphene states are topological ones as at bypass
around the vortex core, electrons and holes acquire nonzero discrete Zak phases $Q$ (see Fig.~\ref{fig1}b).
These Zak phases at high wave numbers $q$, $\vec q = \vec k - \vec K$ form the cyclic group
$Z_8$ with a generator $-{\pi\over 4}$ or ${\pi\over 4}$ for electrons or holes
at sufficiently high wave numbers $q$
as
\begin{eqnarray}
Q = \mp {n\pi\over 4}, \ n=0,1,\ldots \label{topological-charge-values}
\end{eqnarray}
where the $Q$ are called topological charges 
or flavours. To reconcile the presence of
topological charges (\ref{topological-charge-values}) with the electron--hole symmetry  one should
assign zero value to whole graphene topological charge
\begin{eqnarray}
Q_{graphene} = 0. \label{whole-topological-charge}
\end{eqnarray}

Three resonances which emerge in the optical graphene response
$\Re e \ \sigma_{Ohm}$ (the real part of the ohmic contribution to
the optical conductivity) at $\Sigma_{BA}\Sigma_{AB}\
(\Sigma_{AB}\Sigma_{BA})\neq 0$ and are shown in
Fig.~\ref{Majorana-mode-excitation}a, are the direct evidence of
existence for  three topologically non-trivial Majorana-like modes
with different nonzero topological charges, for example, such as
 \begin{eqnarray}
Q_n =\mp {(n+1)\pi\over 4}, \ n=1,2, 3.
\end{eqnarray}
Since a total topological charge 
$Q_{sum}=\sum_{i=1}^{3}Q_i$ of the Majorana-like mode configuration is  non-multiple 
to $2\pi$ and equal to
\begin{eqnarray}
Q_{sum}={\mp 9\pi\over 4}
\label{nonzero-topological-charge}
\end{eqnarray}
the  law of topological-charge conservation prevents the decay
of the Majorana vortical states through topologically trivial electron--hole annihilation.

In the massless case ($\tilde M=0$), the three resonances are degenerated into one
as the comparison of the simulation results for
graphene optical response show for the case with and without mass term
$\Sigma_{BA}\Sigma_{AB}$ ($\Sigma_{AB}\Sigma_{BA}$)
demonstrates (make a comparison between the curves ``1'' and ``2'' in Fig.~\ref{Majorana-mode-excitation}a).
It testifies that the $\Sigma_{BA}\Sigma_{AB}$ ($\Sigma_{AB}\Sigma_{BA}$) describes interaction
of the Majorana modes with different topological charges (flavours) with each other by the following mechanism.
In analogy as electromagnetic interaction pushes out topological defects
(vortex cores) in the neighbourhood of the 1st Landau level, the coupling $\Sigma_{BA}\Sigma_{AB}$
($\Sigma_{AB}\Sigma_{BA}$) creates color energy gap $E_{g,Q_n}(q_n)$, $\vec q_n= \vec k_n - \vec K$ of the
graphene bandstructure that forbides the presence of Majorana modes in the
$K(K')$ point and positions the vortex cores in non-coinciding points
$\vec q_n$, $n=1,2,3$ of the Brillouin zone (see Fig.~\ref{Bandstructure}b).
The $Q$-flavour interactions may proceed before a moment when external electromagnetic field
pushes out the Majorana-like modes into vicinity of the 1st Landau level and visa versa
the flavour Majorana-like vortical modes may be redistributed on the discrete Landau levels
after the impact of the electromagnetic field $\vec A$ on p$_z$ electrons gives rise the flavour interactions.
%
Resonances in the system of coupled oscillators in an external field are called Fano resonances.
The Fano-like resonances  being observed in the real part of $\sigma_{Ohm}$ is
signature  of the graphene vortical states which form a system of three coupled oscillators in an external field.

\subsection{Topological-charge affect on quantum interference }

The phase of the wave function of a graphene charge carrier acquires
an additional Berry phase $\phi_{B}$ upon bypassing any defect,
including a topological one, in the magnetic field $\vec{B}$
of electromagnetic radiation with the vector potential $\vec{A}$.
%
If the defect is not topological one, quantum interference
of charge carriers with different phases would lead to
a change (decrease) in the intensity of the optical response,
and the magnitude of this change depends on the frequency $\omega$ of $\vec{A}$.
The topological charge conservation law prohibits the effect of quenching
of out-of-phase waves that carry non-zero  topological charge,
the last explains the monotonic frequency dependence of the graphene
conductivity at $\tilde{M}\neq 0$ (see Fig.~\ref{Majorana-mode-excitation}).

The observable merging of three Majorana resonances with the total topological charge $Q_{\text{sum}}$
(\ref{nonzero-topological-charge}) into a single one when $\tilde{M}= 0$
(see Figs.~\ref{Bandstructure}b and \ref{Majorana-mode-excitation}a) means that the Majorana modes are degenerate
without the flavour interactions.
The topological charges of all three flavour Majorana modes
that constitute this degenerate mode are located at the same point (the Dirac point $K(K')$).
However, generally speaking, the topological charge $Q_{d}(K)$ of the degenerate
Majorana mode residing at $K(K')$ must vanish in accordance with
the zero value of the graphene topological charge $Q_{{graphene}}$ (\ref{whole-topological-charge}).
%
The topological charge $Q_{d}(K)$ becomes zero under the condition
that a compensating topological countercharge $Q_{0}(K)$,
equal to $Q_{0}(K) = -Q_{\text{sum}}$,
is located at the merging point of the cores of the flavour
Majorana modes (at each point $K(K')$ of reciprocal space), so that
\begin{eqnarray}
 Q_{d}(K)= Q_0(K)+Q_{sum} =0 \ \mbox{for}\ \tilde{M}= 0. \label{antiflavour}
\end{eqnarray}
The  antiflavour $Q_0(K)$ satisfying to the condition (\ref{antiflavour}) must be always resident in
the graphene valley because the requirement (\ref{whole-topological-charge}) holds.

Since in the absence of flavour interaction at $\tilde{M}= 0$
the all flavour vortex cores degenerate into a single flavourless
Majorana-like zero-energy mode, and the electromagnetic field $\vec{A}$
with Abelian statistics is insensitive  to topological charge,
flavourless  vortices as coherent quantum structures
must interfere with each other as quasiparticle excitations with different phases.
Since the Berry phase $\phi_{B}$, acquired by a graphene charge carrier
upon bypassing a topological defect in the magnetic field $\vec{B}$
of electromagnetic radiation, is added to the phase of its wave function,
the interference pattern must oscillate with the variation
of the frequency $\omega$ of $\vec{A}$.
This explains the observed oscillations of $\Re e \, \sigma_{Ohm}(\omega)$ at $\tilde{M}= 0$ (see
the curves ``2'' in Fig.~\ref{Majorana-mode-excitation}).

We note that the oscillation amplitude of the optical conductivity
$\sigma_{Ohm}$ varies with increasing frequency $\omega$ of the applied field $\vec{A}$.
This occurs due to the growth of the Majorana mode feathering
through the formation of new non-overlapping vortex sleeves.
%
Then, one can infer from this that each new sleeve contributes
an additional Berry phase $\phi_{p}$ to the wave function of the Majorana-like vortical state.
To prove this, let us consider how the gradient $\vec{\nabla} E_{q}$
of the energy dispersion $E_{q}$ varies along the vortex sleeve.
The vorticity of graphene charge carrier states is revealed 
as a values drop of $\vec{\nabla} E_{q}$ from $\vec{\nabla} E_{q_1}$ to $\vec{\nabla} E_{q_2}$
with a $90^{\circ}$ rotation of the vector $\vec{\nabla} E_{q}$
along the vortex sleeve in the direction away of the sink (vortex funnel) (see Fig.~\ref{fig1}a) and
\begin{eqnarray}
\vec{ \nabla} E_{q_2} \perp \vec{ \nabla} E_{q_1},\ \
\vec{ \nabla} E_{q_2}\ll \vec{ \nabla} E_{q_1}
\end{eqnarray}
where $\vec{ \nabla}={ \partial \over \partial {\vec q} }$.
The change in the direction of the gradient $\vec{\nabla} E_{q}$
entails a reorientation of the spin of the massless charge
carrier, because chiral symmetry always aligns its spin along the
direction of motion.
According to Fig.~\ref{Bandstructure}a, since electron-hole symmetry admits the energy levels $E_{q_1}$
and $E_{q_2}$ belonging to the electron (hole) and hole (electron) bands and the ${\pi\over 2}$-angle turn is equivalent
to ${\pi\over 6}$ due to the hexagonal symmetry,
respectively, the quantum exchange can perform a transfer
of the electron (hole) from a valley $K_{A_1}(K'_{B_1})$
into a valley $K_{A_2}(K'_{B_2})$ through both the Dirac $K'_{B}(K_{A})$ and  $M$ points
at transition between the energy levels $E_{q_1}$ and $E_{q_2}$ under an external affect.
The process of  alternating transition of an electron (hole)
from one trigonal sublattice $A(B)$ to another $B(A)$
and back to $A(B)$ must be accompanied by a periodic change
in  spin direction. Such a process is called precession.
The precession process is analogous to the bending of the trajectory
of an electric charge in a magnetic field $\vec B$ due to the electron--hole symmetry.
Therefore, just as in the case of bypassing a topological defect in a magnetic field $\vec B$, the graphene
charge carriers precessing along the $j$-th sleeve  of the vortex acquire
an additional Berry phase $\phi_{p,j}$, $j=1,2, \ldots$.
%
Since the sleeves do not coincide and their widths vary, $\phi_{p,i}\neq
\phi_{p,j}$, $i\neq j$. The fact that with increasing frequency $\omega$ the Berry
phase becomes different, taking the value of $\phi_{B}+\sum_{k=1}^j
\phi_{p,k}$, explains the observed aperiodic variation
of the interference pattern with increasing $\omega$ (see Fig.~\ref{Majorana-mode-excitation}).

We note (see the curves ``2'' in Fig.~\ref{Majorana-mode-excitation}a) that  $\sigma_{Ohm}$ value oscillates
in the vicinity of half the graphene conductivity quantum of $0.25 G$ (the conductivity
quantum or minimal graphene conductivity is equal to $0.5 G$, $G={e^2\over h}$).
The value of $0.25 G$ is the optical conductivity of the topologically trivial pseudo-Dirac model.
Such behaviour confirms the assumption on the existence of a 4th flavour
$Q_0$ not associated with any physical Majorana-like mode.
Thus, the oscillations of graphene optical conductivity at $\tilde M =0$, arising
due to the existence of the topological anticharge $Q_0=-{\pi\over 4}$ in our model, disappear
due to the removal of degeneracy of the Majorana-like vortex cores by
$\Sigma_{BA}\Sigma_{AB}\neq 0$ ($\Sigma_{AB}\Sigma_{BA}\neq 0$).
The value $Q(q_i)$, $i=0,\ldots,3$ of the topological charge being
pushed out from the graphene valley to the point $\vec k_i=\vec q_i +
\vec K$ of the Brillouin zone tends to $Q_i(0)$ at an infinitesimally small
Majorana term, $\tilde M\to 0$.
It means that according to the expression (\ref{antiflavour}) the total topological charge $Q_{\text{total}}$
of the flavour Majorana-like modes tends to the following value:
\begin{eqnarray}
Q_{total}=\sum_{i=0}^3 Q(q_i) \to \sum_{i=0}^3 Q_i(K)= Q_{d}(K) =0 \ \mbox{for}\ \tilde{M}\to 0 \label{antiflavour1}
\end{eqnarray}
and, correspondingly, the topological charges $Q_i(K)$, $i=0,\ldots,3$ accumulated in $K(K')$ are completely
pushed out from the graphene valley.
%
Since the topological antidefect (antiflavour $Q_0$) can coexist with the
topological defect (flavour $Q_i$, $i=1,2,3$) only at the Dirac
point $K(K')$ due to the chiral symmetry, the anti-Majorana state
with the topological charge $Q(q_0)$ becomes unstable or
unphysical and, correspondingly, the topological charge conservation
law must prevent the quantum interference of the physical Majorana-like states with the
flavour $Q(q_i)$, $i=1,2,3$ as confirmed by the
non-oscillating behaviour of the calculated optical graphene conductivity
at $\Sigma_{BA}\Sigma_{AB}\neq 0$ ($\Sigma_{AB}\Sigma_{BA}\neq 0$).
However, the behaviour of $\Re e \ \sigma_{Ohm}$ becomes kink-like at high frequencies $\omega$
when the flavour Dirac cone bands begin to overlap and then mix (see the curve ``1'' in
Fig.~\ref{Majorana-mode-excitation}a, left).
It is evidence of the fact that the presence of antiflavour
electron (flavour hole) Majorana vortical states can give rise to
this quenching of the Majorana waves at high energies ($\hbar \omega$).

Further, let us explore the origin of such flavour electron (antiflavour
hole) Majorana-like vortical states to which the antiflavour (flavour) topological defects are admixed.

\subsection{Chiral-symmetry recovery}

\begin{figure*}[hbtp]
\begin{center}
(a) \ \ \ \includegraphics[width=5cm,angle=0]{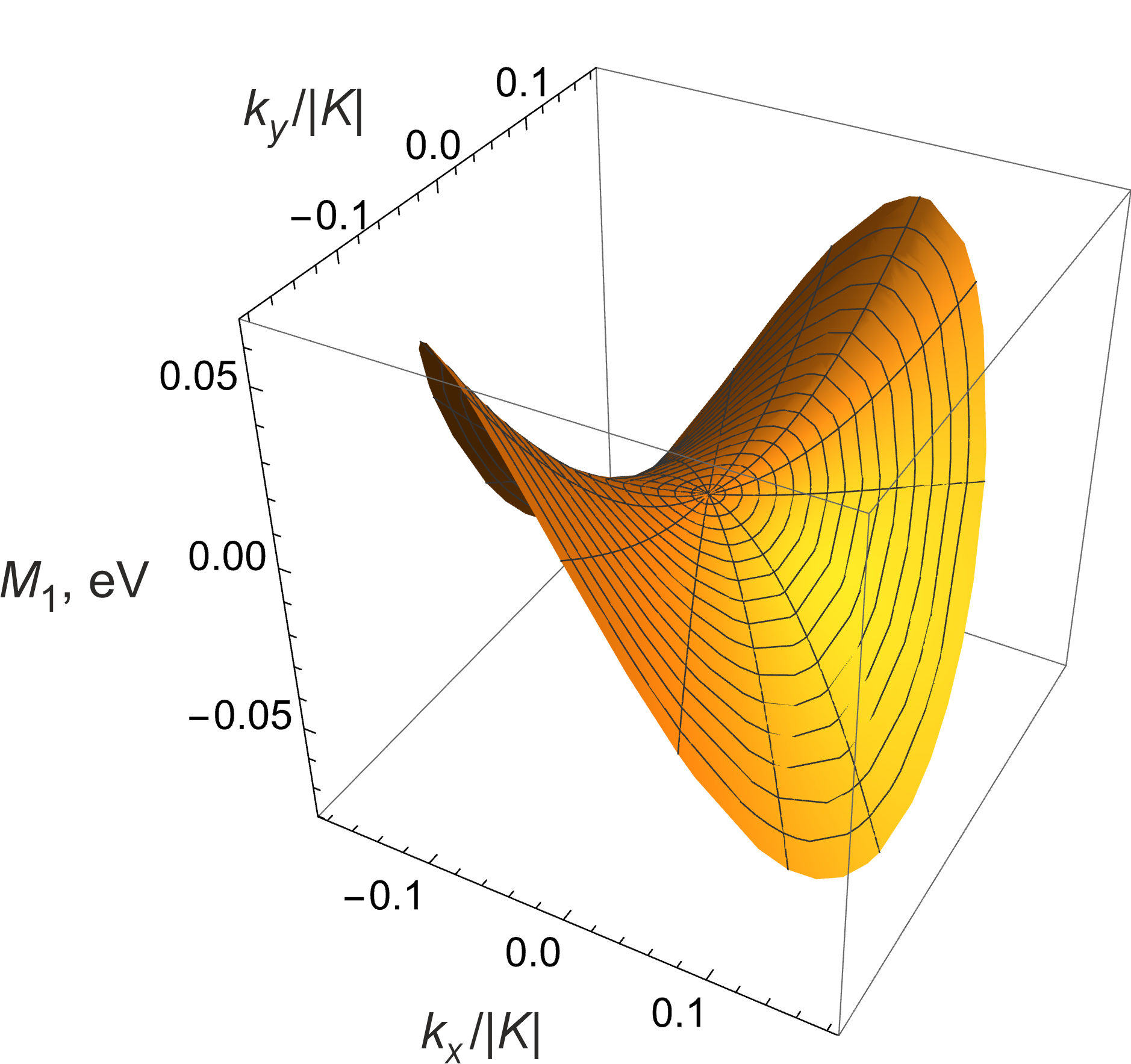}
 \includegraphics[width=5cm,angle=0]{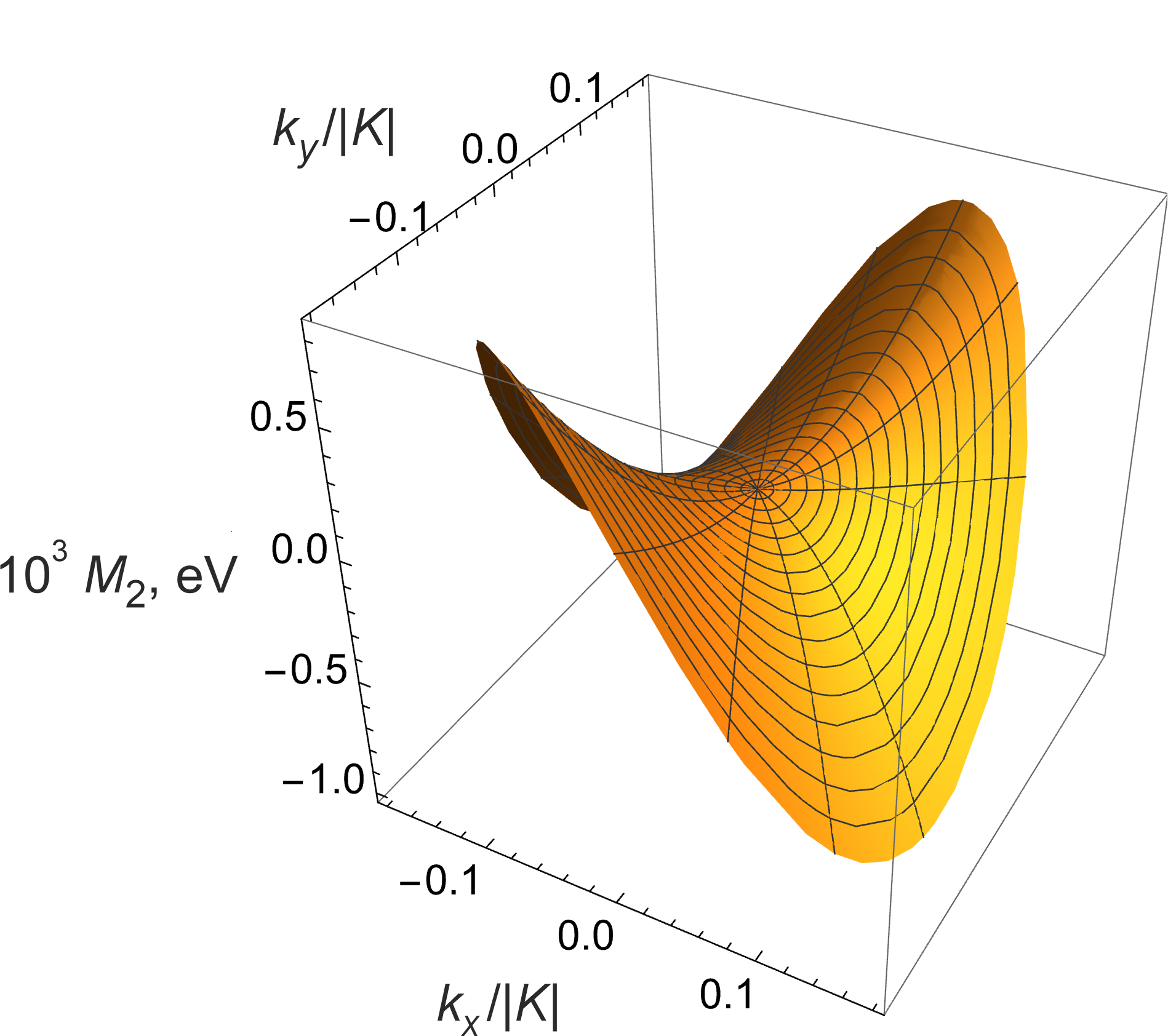}\\
(b) \ \ \ \includegraphics[width=5cm,angle=0]{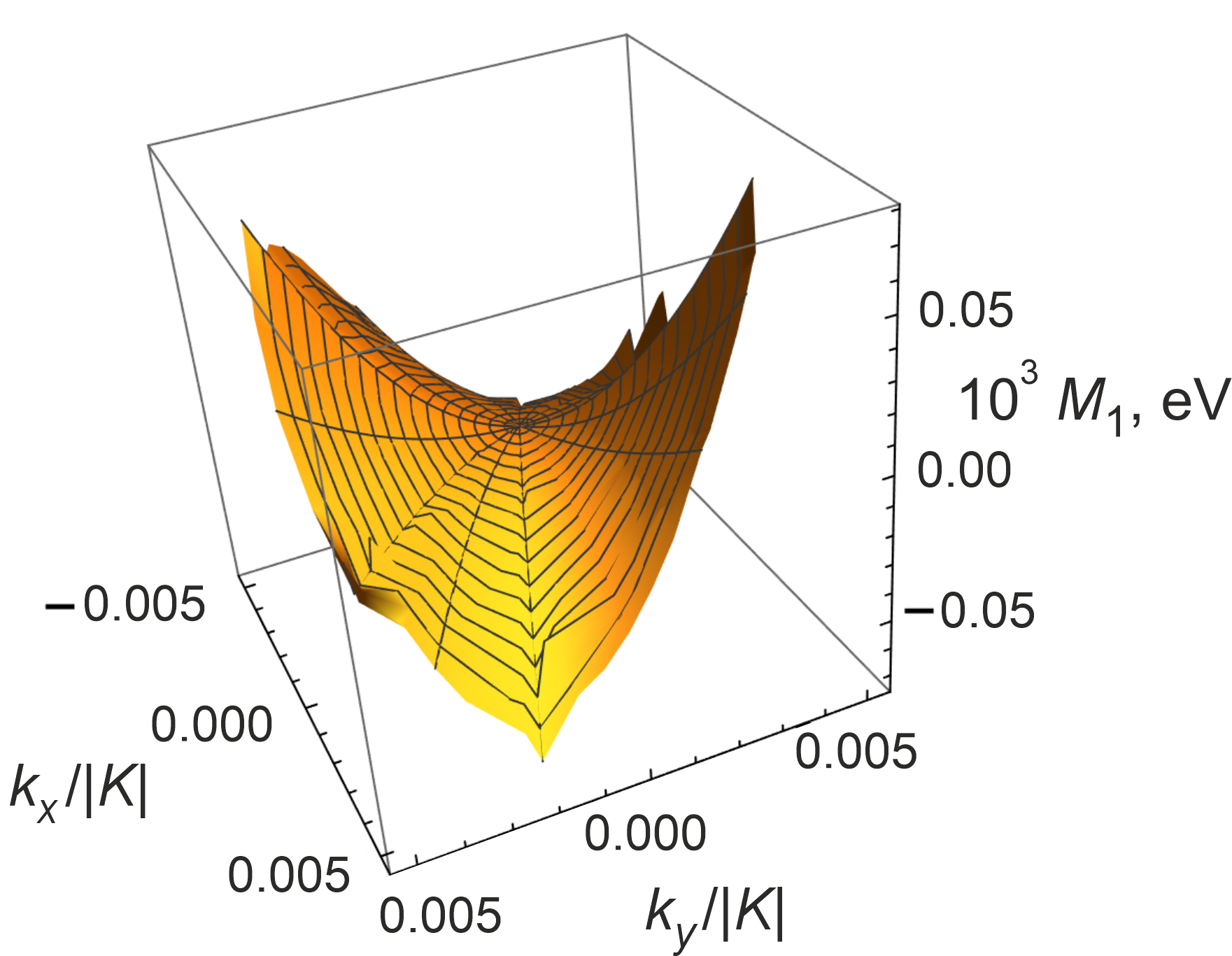}
\includegraphics[width=5cm,angle=0]{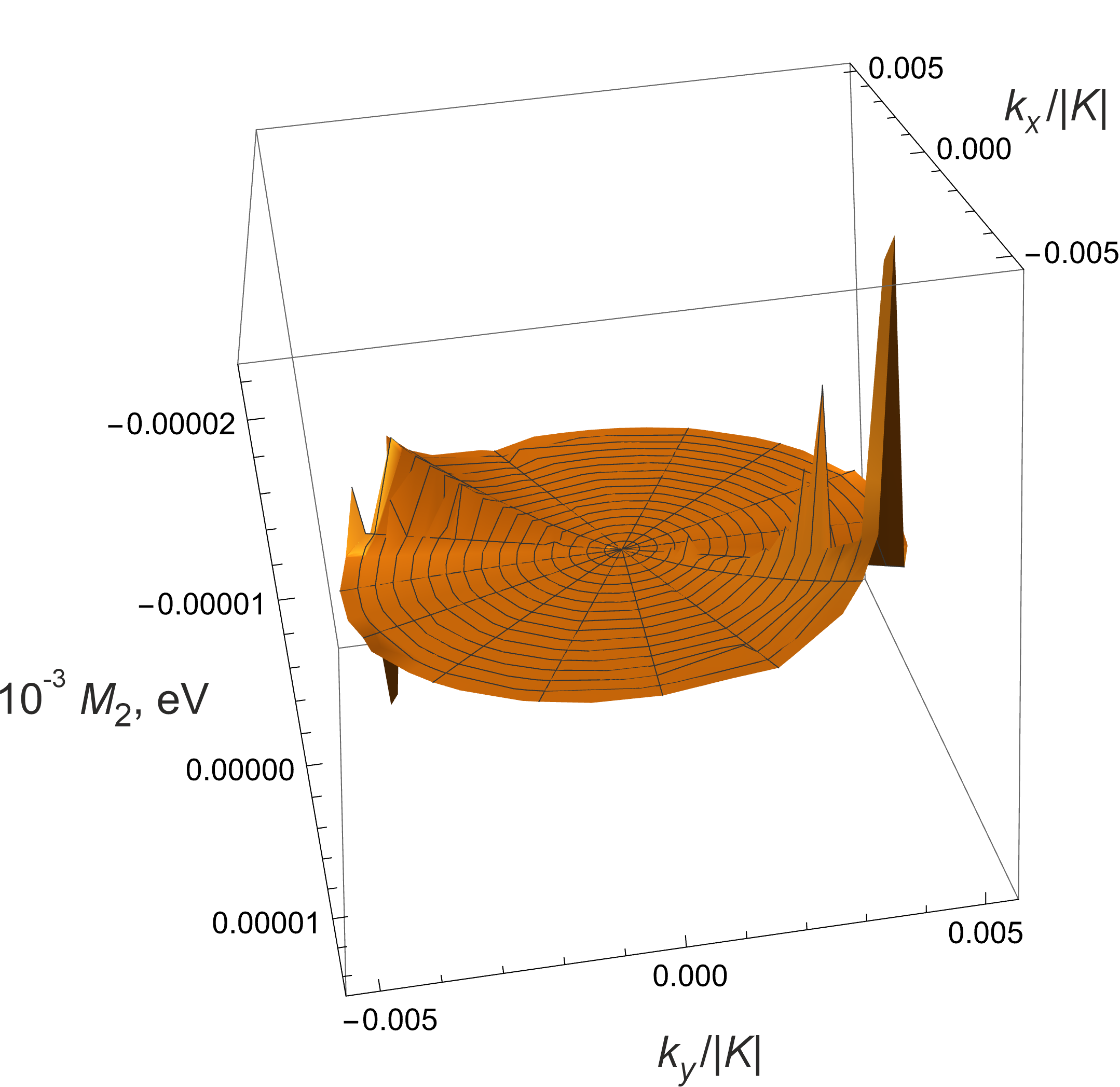}\\
(c) \ \ \ \includegraphics[width=5cm,angle=0]{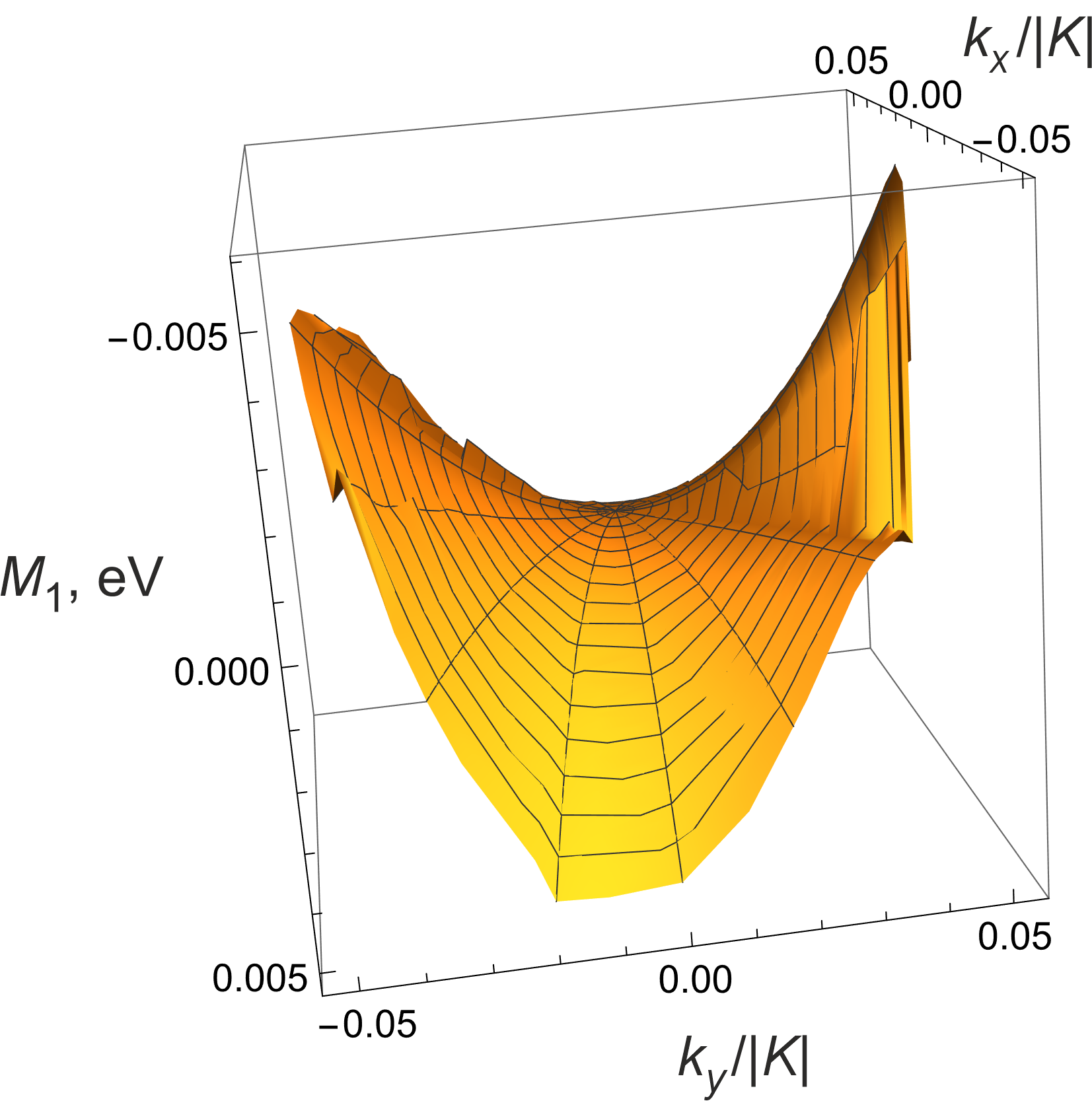}
\includegraphics[width=5cm,angle=0]{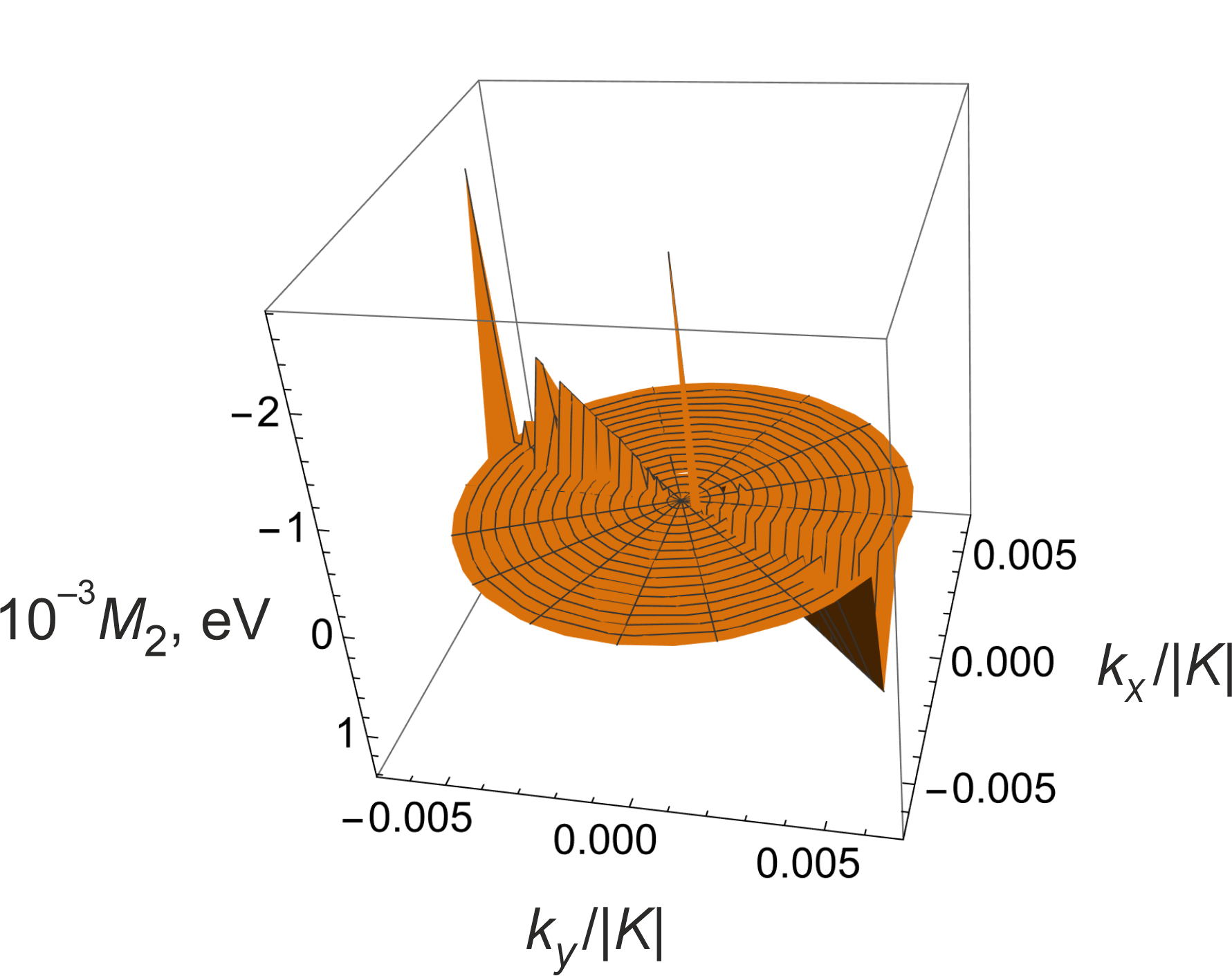}\\
(d) \ \ \includegraphics[width=5cm,angle=0]{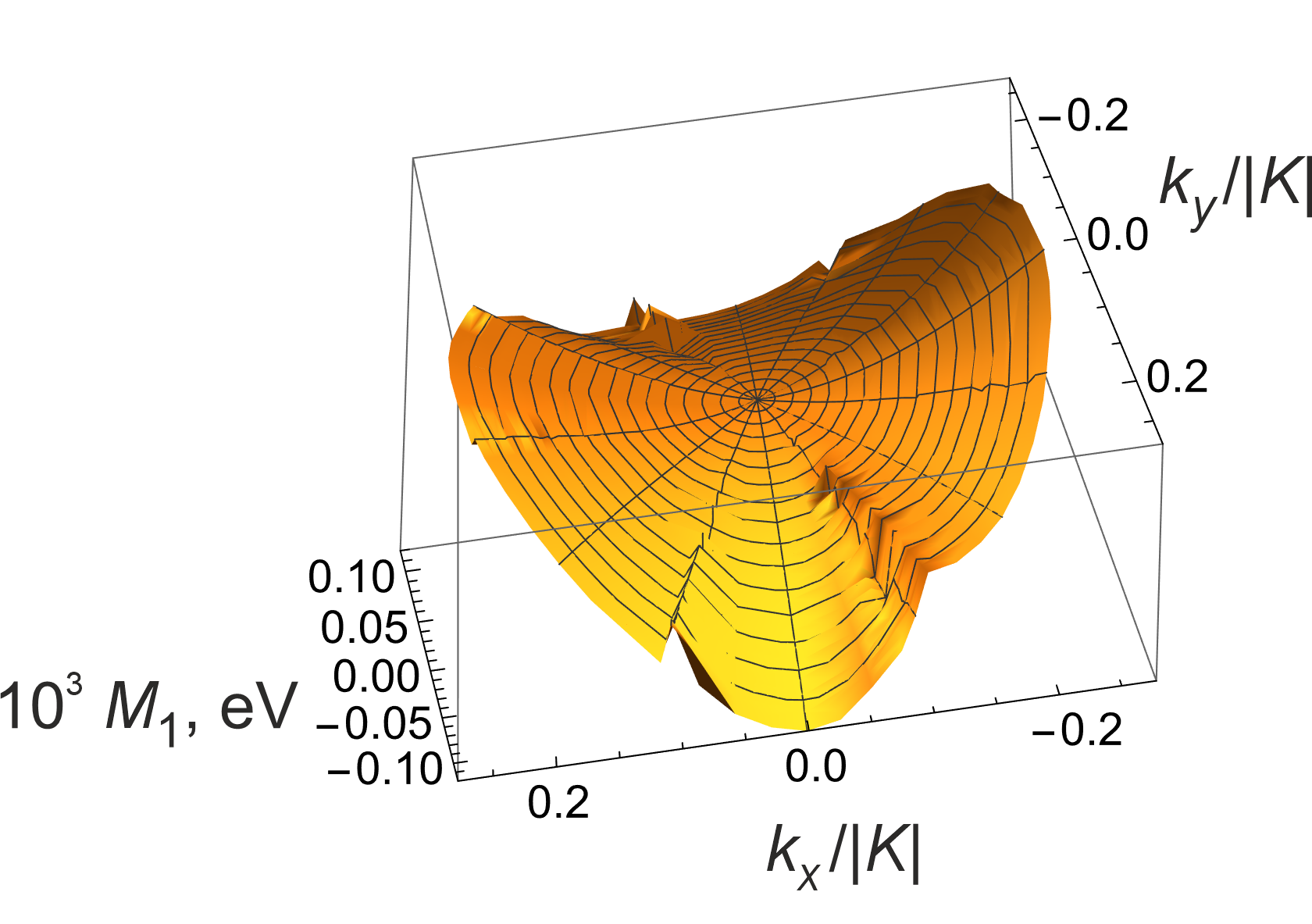}
\includegraphics[width=5cm,angle=0]{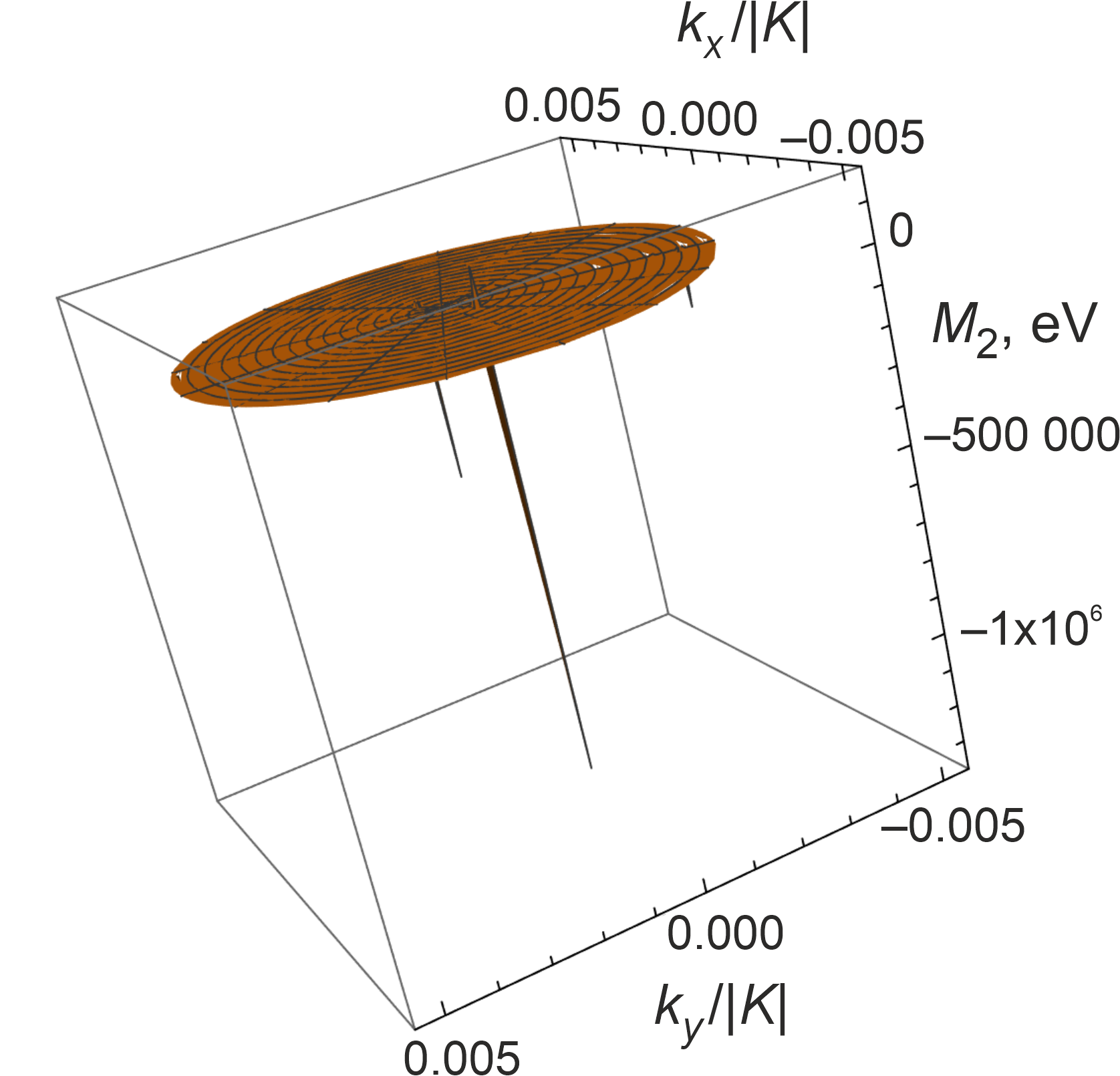}
\end{center}
\caption{Two eigenvalues, $M_1$ and $M_2$, of the Majorana mass term at zero (a) and non-zero (b--d) gauge fields,
without spin-valley current coupling (b), and with small (c) and large (d) spin-valley current coupling.}
 \label{Mass-term-eigenvalues}
\end{figure*}

Under an assumption of unchanged phases of the p$_z$ electron wave functions (zero gauge fields when
 the interaction ($2\times 2$)-matrices $\Delta _{AB}$ ($\Delta_{BA}$)=1), two $\Sigma_{BA}\Sigma_{AB}$
($\Sigma_{AB}\Sigma_{BA}$) eigenvalues $M_i^{(0)}(q)$ ($-M_i^{(0)}(q)$), $i=1,2$ depending on
$\vec q =\vec k - \vec K_A $ ($\vec q =- \vec k - \vec K_B $), are functions with saddle-type singular points
residing in the graphene valleys; moreover, outside the valley one of the eigenvalues
exceeds the other by two orders of magnitude (see Fig.~\ref{Mass-term-eigenvalues}a).

%
The electron--hole symmetry of massless graphene charge
carriers, forbidding a change in their chirality, turns an electron
(hole) into a hole (electron) upon reversal of the direction of the vector $\vec q $, $\vec q \to -\vec q$.
According to Fig.~\ref{Mass-term-eigenvalues}a, outside the
Dirac point, due to  preservation of sign under the substitution $\vec q \to -\vec
q$, the mass eigenvalue $M_i^{(0)}(q)$, $i=1,2$ violates chiral
symmetry and, correspondingly, electron--hole pairs annihilate, destroying the feathering  of the vortex cores
upon the transition from one ``wing'' (``antiwing'') of the saddle to the opposite one.
Thus, in the absence of gauge fields, the chiral
Majorana modes outside the valley are unstable and can exist only at zero energy.

When the gauge fields are non-zero, the only one ($M_1(q)$) of
 two eigenvalues $M_1(q)$ and $M_2(q)$ of the Majorana-like term
$\Sigma_{BA}\Sigma_{AB}$ ($\Sigma_{AB}\Sigma_{BA}$) remains
a saddle, and the span of its ``wings'' sharply decreases (see Fig.~\ref{Mass-term-eigenvalues}b):
\begin{eqnarray}
M_1(q) \ll M_1^{(0)} (q), q>0
\end{eqnarray}
and, correspondingly, the feathering of one of the two possible vortex pairs always annihilates.
%
Another eigenvalue $M_2(q)$ flattens (see Fig.~\ref{Mass-term-eigenvalues}b)
so that the ``antiwings'' of the saddle vanish, and three mass resonances $M_{r}(q_i)$,
$i=1,2,3$ grow on one of the rudiments of the ``wings''.
Since the vortex-core energy $E_0$ is zero, the mass resonances give rise to a flavoured energy gap $E_{g}
(q_i)$, $i=1,2,3$ of the graphene bandstructure as
\begin{eqnarray}
E_{g} (q_i) = E_0+ M_{r}(q_i), \ i=1,2,3
\end{eqnarray}
and the three flavour Dirac cones at the points $q_i$, $i=1,2,3$
arise (see Fig.~\ref{Bandstructure}b). 
The topological-charge-conservation law (\ref{antiflavour1}) admits only the simultaneous
coexistence of the three mass resonances and, in this sense,
the three mass resonances give rise to an integral indivisible system
of three coupled Majorana-like modes which reside in different
non-overlapping vicinities of the Dirac point.
This is the cause of the emergence of Fano resonances in the optical response
(see the curve ``1'' in Fig.~\ref{Majorana-mode-excitation}a).
Thus, the topological vortical resonances with three different
flavours $Q_i$ acquire a mass $E_{g,Q_i} (q_i) = M_{r}(q_i)$,
$i=1,2,3$ when the mass term is coupled to the gauge fields. Now, we  examine the stability of these states.

Let us numerically prove the coexistence of the Majorana-like modes with
the anti-flavour ones in the following way.
The precessing spin $\vec \sigma^{AB} (\vec \sigma^{BA})$ is no longer orthogonal to the valley
currents $\vec K^{BA}_B+\vec K^{BA}_A\ (\vec K^{AB}_B+\vec K^{AB}_A)$ because
\begin{eqnarray}
\vec \sigma^{AB} \cdot \left(\vec K^{BA}_B+\vec K^{BA}_A \label{precession}
\right)\neq 0.
\end{eqnarray}
The  spin precession (\ref{precession}) give rise nonzero values of the
added mass with  $m_{precession}$ (\ref{m-precession}) entering the mass term $\tilde M$
(\ref{orbital-valley-currents-coupling1}),(\ref{spin-valley-coupling}).
As one can see in Fig.~\ref{Mass-term-eigenvalues}c, the effect of a nonzero value of $m_{\text{precession}}$
of the order of 0.0001 consists in a further flattening of one ($M_2(q)$) of the eigenvalues of
the Majorana-like term with the replacement of one of the ``wings'' of the saddle by an ``antiwing''.
Thus, the spin--valley-current interaction restores the chiral symmetry of graphene charge carriers.
The three resonances are chirality anomalies.

At large $m_{\text{precession}}$ of the order of 0.001, the eigenvalue
$M_2(q)$ vanishes for all $q$ except four wave numbers
$q_i$, $i=0,1,2,3$ at which chirality anomalies remain in the form
of three resonances with the flavours $Q(q_i) = {(i+1)\pi\over 4}, \
i=1,2, 3$ and one antiresonance with the antiflavour
$Q(q_0)=-{\pi\over 4}$ (see Fig.~\ref{Mass-term-eigenvalues}d).
%
It is important that the zero-energy
Majorana-like mode itself with negative (positive) mass $Q(q_0)$,
$Q(q_0)<0$ ($Q(q_0)>0$) as a mass of negatively (positively)
charged graphene electrons (holes) has no physical meaning.

However, due to the electron--hole symmetry there always exists
a superposition (mixture) of Majorana anti-pairs of vortex cores and
chiral pairs of developed vortices with masses $M_1(q)$ and $M_2(q)$, respectively.
Therefore, the $Q(q_0)$ can always be assigned
to topological antidefects with a saddle-like dependence of the mass on
$\vec q$ in the vicinity of the $K'(K)$ antivalley.

Thus, the flavour interaction $\Sigma_{BA}\Sigma_{AB}$ ($\Sigma_{AB}\Sigma_{BA}$) render the vortical states
massive as  LH  electron (RH hole) stable-vortices core with the flavour $Q_n$ (antiflavour $-Q_n$), $n=1,2,3$
acquire a mass 
$M_r(q_i)>0$ ($-M_r(q_i)<0$), $i=1,2,3$  and the antimass $M_r(q_0)<0$ ($-M_r(q_0)>0$)  is
attributed to the RH electron (LH hole) unstable-vortex core with antiflavour (flavour) $Q_0$.

The interaction $m_{\text{precession}}$, mediated by the precession
caused by the quantum exchange, renders null the ability of the
flavour interaction $\Sigma_{BA}\Sigma_{AB}$
($\Sigma_{AB}\Sigma_{BA}$) to violate the chirality outside the graphene
valleys for the LH electron (RH hole) vortex cores only.
%
Then, our theory predicts that the flavoured vortex resonances will emerge at
sufficiently low energies, when the law of topological-charge conservation
precludes quantum interference through the annihilation of feathering with the Majorana zero-energy modes
attributed to the mass-term eigenvalue $M_1(q)$.
%
At high energies, both stable and unstable vortices develop, with topological charges of opposite
sign; their total zero topological charge admits mutual quantum interference.

Now, one is able to construct associations between the concepts and
terms of the Majorana scenario of admixing a sterile RH
mass neutrino to three active mass LH neutrinos and the concepts and
terms of the Majorana-like graphene physics.

In the absence of an electromagnetic field and, correspondingly, without
electron--hole pairs, there exist three Majorana-like LH modes
with positive electron (negative hole) masses $M_r(q_i)$ ($-
M_r(q_i)$), $i=1,2,3$ residing in the graphene sublattice
$A$ ($B$) and one RH Majorana-like antimode with a mass $M_r(q_0)$ of the opposite sign.
The Majorana-like antimode can be attributed to the other sublattice $B$ ($A$) and then it, as
a real antimode, would be admixed to the real Majorana-like modes,
remaining just as non-interacting (sterile) due to residing on the other sublattice because of chiral symmetry.

In a similar way, as in the flavoured graphene interaction, when neutrinos
and electromagnetic plasma are decoupled, the three types of LH neutrinos (active neutrinos) are
revealed  explicitly in weak interactions and there also exists
an RH neutrino of the 4th type (sterile neutrino, sterile RH lepton),
which can be admixed without an additional resonance.

%
At low energies, when coupling takes place between
electromagnetic fields and  Majorana-like graphene modes of one
sublattice $A$ ($B$) only, without the antimode from the another
sublattice $B$ ($A$), the real flavour Majorana-like modes are exhibited through the Fano resonances.
At higher energies, when the unstable high-energy antiflavour Majorana-like vortical state develops,
quantum interference of all Majorana-like vortical states occurs
because of the overlap of all flavour and antiflavour graphene Dirac cones with each other.
%
Similarly, in weak interactions, since the three active flavour
neutrinos possess physical masses, the sterile neutrino,
being periodically admixed, firstly, modifies the neutrino mass
spectrum and, secondly, at high energies, by decolouring
the neutrino beam, causes its components to interfere with
each other, which gives rise to the three-flavour neutrino oscillations.

\section{Discussion and conclusion}
%
So, according to the quasi-relativistic quantum-field theory of graphene Majorana-like fermions,
zero-energy pseudo-Majorana modes reside in the Dirac point $K(K')$ of the graphene
Brillouin zone as topologically nontrivial defects.
Our approach elucidates
 physical meaning of the topological charge, the flavour of leptons
is associated with the topological charge $Q_i$, $i=0,1, \ldots$ of
the vortex core.

The three Majorana-like flavour modes with repulsive interaction emerge in the graphene optical conductivity.
 The theoretical prediction of flavour Fano resonance  with  maximum 
 of $4.74$~eV ($5.5 \times 10^4$~K)
is in excellent agreement with maximum 
of 4.62~eV for an experimental ultraviolet 
graphene Fano interference which was registered earlier in 
\cite{Mak-et-al2011}. 
The predicted resonances ``F1'' and ``F2'' being confirmed by the presence of the experimentally
observed asymmetric UV spectrum line in the optical
conductivity of graphene ensure that its value exceeds
that predicted by the pseudo-Dirac model of graphene.
For comparison, in other papers,
the asymmetric peak ``F2'' is explained only phenomenologically as the existence of a weak coupling between
photon scattering on Dirac-type graphene fermions with a
continuous spectrum and the excitation followed by rapid decay
of excitons with a discrete spectrum on flat regions in the vicinity
of the $M$ point of the Brillouin zone
%
(see \cite{Mak-et-al2012}  
and references therein).
The ${ab\text{-}initio}$ Green's-function (GW) calculation, by fitting
the so-called Fano resonance parameters, predicts the ``F2''
780~meV bandwidth (exciton lifetime of the order of 0.5~fs) which
is experimentally observed; but, the predicted excitonic 5.2~eV
peak is blue-shifted relative to the experimental one.
%
Conversely, the ${ab\text{-}initio}$ GW--Bethe--Salpeter (GWBS) calculation,
by fitting the Fano resonance parameters, predicts a close
proximity of the excitonic peak to
the ``F2'' peak (5.02~eV); but, the excitonic-peak bandwidth of the order of 200~meV 
corresponds to a slow excitonic decay (exciton lifetime of the order of 2.5~fs), which contradicts the excitonic
instability in graphene.
In Ref.~\cite{Kuzian-et-al2025},
%
%
without elucidating the origin of the weak coupling between the two oscillators,
a discrete level of a three-dimensional system with a quantum well with the attractive
singular $\delta$-potential type to the plane is used,
phenomenologically incorporated into graphene, which yields a broadening
equivalent to a rather slow decay, albeit faster
in comparison with the excitonic one.

%
Both two eigenvalues $M_1(q)$ and $M_2(q)$ of the Majorana mass term $\tilde M$ vanish at the Dirac
points, but in the neighborhood of the Dirac point, only one eigenvalue ($M_2(q)$) remains zero.
The chirality of the   topologically protected Majorana-like mode is violated by four resonances of the second
mass-term eigenvalue $M_2(q)$
.
This chirality anomalies causes the electron (hole) conical Dirac bands
to diverge
.
The four resonances can be considered as elementary excitations (quanta)
of a certain flavour field.
It has been proved that the degenerate zero-energy Majorana-like mode at the Dirac point $K(K')$ possesses
zero total topological charge $Q=\sum_{n=0}^4 Q_i$.
Under the condition that the cores of all vortical states reside in the valley
$K(K')$, there are no prohibitions from the law of
topological-charge conservation on the existence of quantum
interference of the vortex featherings as the cause of the observed oscillations
of the optical graphene conductivity  in Fig.~\ref{Majorana-mode-excitation}a.

%
Since the mass term opens a flavoured gap $E_{g,Q_i}(q_i)$, $i=1,\ldots, 4$ in the
graphene bandstructure, the zero-energy Majorana-like modes become massive.
Assigning spin and the corresponding topological charge to the flavour-field quanta, one obtains four
flavour leptons.
The three Majorana-like LH (RH) modes of them possess positive (negative) masses $M(q_i)=E_{g,Q_i}(q_i)>0$,
$i=1,\ldots, 3$ and participate in physical repulsive interactions with the preservation of chiral symmetry.

%
Since all vortical Majorana-like states are a superposition of the eigenstates of the Majorana mass term
$\tilde {M}$, the unstable RH (LH) vortical Majorana-like state with the alternating-sign eigenvalue
$M_1(q)$ is always admixed to the three-flavour chiral model outside the Dirac point, and,
correspondingly, the mass $M(q_0)=E_{g,Q_0}(q_0)<0$
$(M(q_0)=E_{g,Q_0}(q_0)>0)$  is always an attribute
of the fourth non-chiral lepton, the interaction with which is suppressed by CP symmetry.

%
In our theory, the Majorana-like mixture of LH-chiral and  RH-non-chiral vortical states
when holding total zero topological charge
is revealed through the quantum interference as three-flavour
oscillations because the law of topological-charge conservation
admits the redistribution of the energy among the flavoured constituents
of the flavourless composition.

We offer a Majorana scenario of three-flavour mass-neutrino
oscillations by drawing parallels and interpreting
the phenomenological mixing of LH active neutrinos with an RH sterile
lepton as the formation of a bandstructure gap by means of
the interaction of nonzero topological charges of Majorana-like modes.
A necessary and sufficient condition for this scenario is the
existence of such Majorana modes with topological charges
of opposite sign which reside on different sublattices.
The predicted aperiodic variation of the optical graphene
response for the mixed Majorana-like state explains the three-flavour neutrino oscillations.
Within the framework of the proposed toy model, these oscillations can be interpreted as
the existence of a sterile $M(q_0)$-mass RH antiflavour neutrino,
periodically admixed to the three chiral LH active neutrinos with real physical flavours.
The all flavours are different topological charges $Q_n$, $\sum_n Q_n=0$.


\begin{thebibliography}{99}
\bibitem{Esteban2020}
I. Esteban, M.C. Gonzalez-Garcia, M. Maltoni, T. Schwetze, A. Zhou.
The fate of hints: updated global analysis of
three-flavour neutrino oscillations. JHEP. {\bf 2020}, 178 (2020).

\bibitem{Escudero2019} M. Escudero. Neutrino decoupling beyond the Standard Model: CMB constraints on
the Dark Matter mass with a fast and precise $N_{eff}$ evaluation.
J. Cosmol. Astropart. Phys. {\bf 2019}, 007 (2019).

\bibitem{Binder2023} T. Binder, M.~Garny, J.~Heisig, S.~Lederer, K.~Urban. Excited bound states and
their role in dark matter production. Phys. Rev. D. {\bf 108},
095030 (2023).

\bibitem{Esteban2019} I. Esteban, M.C. Gonzalez-Garcia, A. Hernandez-Cabezudo. Global analysis of
three-flavour neutrino oscillations: synergies and tensions in the
determination of $\theta_{23}$, $\delta_{CP}$, and the mass
ordering. JHEP. {\bf 2019}, 106 (2019).


\bibitem{Vries2025} J. de Vries, H.K. Dreiner, J. Groot, J.Y. G\"unther, Z.S. Wang.
Probing light sterile neutrinos in left-right symmetric models
with displaced vertices and neutrinoless double beta decay.
J.~High Energ. Phys. {\bf 2025}, 7 (2025).

\bibitem{Chaturvedi2024} K. Chaturvedi. Study of sterile neutrino contribution to
neutrinoless double beta decay. SCIREA J. Phys. {\bf 9}, 130
(2024).

\bibitem{Adams2026} C. Adams {\it et al.} (NEXT Collaboration). The NEXT-100 Detector. Eur.~Phys.~J.~C.
{\bf 86}, 114 (2026).

\bibitem{Acero-et-al2024} M.A. Acero  et al. White paper on light sterile neutrino searches and related
phenomenology. J. Phys. G Nucl. Part. Phys. {\bf 51}, 120501 (2024).

\bibitem{Gagliardi-et-al2025} G. Gagliardi, C. K\"{o}hler, T. Lasserre, S. Mohanty, X. Stribl, The KATRIN Collaboration.
Sterile-neutrino search based on 259 days of KATRIN data. Nature. {\bf 648}, 70 (2025).

\bibitem{Ren2023} H.-Y. Ren, Y.-N. Ren, Q. Zheng, J.-Q. He, L. He.
Electron-electron interaction and correlation-induced two density
waves with different Fermi velocities in graphene quantum dots.
Phys. Rev. B. {\bf 108}, L081408 (2023).

\bibitem{8-braid}
H. Grushevskaya, G. Krylov.
{ Two-Dimensional Braiding of Two-Dimensional Majorana
Fermions : Manifestation in Band Structure of Graphene}. Int J. Nonlin. Phen. Compl. Sys.  {\bf 22}, 41 (2019).

\bibitem{Zak1989}
J. Zak. Berry's phase for energy bands in solids. Phys. Rev. Lett. {\bf 62}, 2747 (1989).

\bibitem{Vanderbilt2018}
D. Vanderbilt, Berry Phases in Electronic Structure Theory: Electric Polarization, Orbital Magnetization and
Topological Insulators, (Cambridge University Press, Cambridge, 2018).

\bibitem{Huang-et-al2025}
Y. Huang, Y. Liu, M. Wang, X. Chen, H. Han, A. Yu, G.P. Wang. Topological transition of
Pancharatnam--Berry phase in a nonlocal twisted bilayer metasurface. Scientific Reports. {\bf 15}, 11182 (2025).

\bibitem{Griffiths-Schroeter}
D.J.  Griffiths, D.F. Schroeter. {\it Introduction to Quantum Mechanics}, third edition.
(Cambridge University Press, Cambridge,
2018)

\bibitem{our-symmetry2020}
H.~ Grushevskaya, G.~Krylov. Vortex Dynamics of Charge Carriers in the Quasi-Relativistic Graphene
Model: High-Energy $\vec{k}\cdot \vec{p}$ Approximation. { Symmetry}. {\bf 12}, 261 (2020).

\bibitem{myJNPCS2025poles} H.V. Grushevskaya, G.G. Krylov. Magneto-optical anomalies and Majorana-like fermion
interactions in graphene. Int J. Nonlin. Phen. Compl. Sys.  {\bf 28}, 144 (2025).

\bibitem{Taylor2016} H.V. Grushevskaya, G.G. Krylov.
Chapter 9. Electronic Structure and
Transport in Graphene: QuasiRelativistic Dirac-Hartree-Fock Self-Consistent Field Approximation.
In: {\it Graphene Science Handbook: Electrical and Optical
Properties}. Vol. 3. Chapter 9. Eds. M.
Aliofkhazraei {\it et al.}
(Taylor and Francis Group, CRC Press, USA, UK, 2016). Pp.117-132.

\bibitem{myNPCS18-2015} H.V. Grushevskaya, G. Krylov.
Semimetals with Fermi Velocity Affected by Exchange
Interactions: Two Dimensional Majorana Charge Carriers.
J. Nonlin. Phenom. in Complex Sys. {\bf 18}, no. 2, 266-283
(2015).

\bibitem{NPCS18-2015GrushevskayaKrylovGaisyonokSerow} H.V. Grushevskaya,
G. Krylov, V.A. Gaisyonok, D.V. Serow. Symmetry
of Model N = 3 for Graphene with Charged Pseudo-Excitons.
Int. J. Nonlin. Phenom. in Complex Sys. {\bf 18}, no. 1, 81-98
(2015).

\bibitem{Mak-et-al2011}
K.F. Mak, J. Shan, T.F. Heinz. Seeing many-body effects in single- and few-layer graphene:
Observation of two-dimensional saddle-point excitons. Phys. Rev. Lett. {\bf 106}, 046401 (2011).

\bibitem{Mak-et-al2012}
 K.F. Mak, L. Ju, F. Wang, T.F. Heinz. Optical spectroscopy of graphene: From the far infrared to the ultraviolet.
  Solid State Com.  {\bf 152}, 1341 (2012)

\bibitem{Kuzian-et-al2025} R. O. Kuzian, D. V. Efremov, E. E. Krasovskii. Fano physics behind the N-resonance in graphene.
 Phys. Rev. Res. { \bf 7} 013180 (2025.).

\end{thebibliography}
\end{document}